\documentclass[graybox]{svmult}

\usepackage{mathptmx}       
\usepackage{helvet}         
\usepackage{courier}        
\usepackage{type1cm}        
\usepackage{makeidx}         
\usepackage{graphicx}        
\usepackage{multicol}        
\usepackage[bottom]{footmisc}

\usepackage{color}
\usepackage{url}
\usepackage[rightcaption]{sidecap}
\newcommand{\Tr}{\mathrm{Tr}}
\newcommand{\sgn}{\mathrm{sgn}}

\usepackage[normalem]{ulem}  

\begin{document}

\title*{Inverse magnetic catalysis in field theory and gauge-gravity duality}
\titlerunning{Inverse magnetic catalysis}

\author{Florian Preis, Anton Rebhan and Andreas Schmitt}

\institute{Florian Preis, Anton Rebhan, Andreas Schmitt\at Institut f\"ur Theoretische Physik, Technische Universit\"at Wien, 1040 Vienna, Austria \\
\email{fpreis,rebhana,aschmitt@hep.itp.tuwien.ac.at}}

\maketitle

\abstract{We investigate the surface of the chiral phase transition in the three-dimensional parameter space of temperature, baryon 
chemical potential and magnetic field in two different approaches, the field-theoretical Nambu--Jona-Lasinio (NJL) model and the holographic 
Sakai--Sugimoto model. 
The latter is a top--down approach to a gravity dual of QCD with an asymptotically large number of colors and becomes, in a certain limit, 
dual to an NJL--like model. 
Our main observation is that, at nonzero chemical potential, a magnetic field can {\it restore} chiral symmetry, in apparent contrast 
to the phenomenon of magnetic catalysis. This ``inverse magnetic catalysis'' occurs in the Sakai-Sugimoto model and, for sufficiently large 
coupling, in the NJL model and is related to the 
physics of the lowest Landau level. While in most parts our discussion is a pedagogical review of previously 
published results, we include new analytical results for the NJL approach and a thorough comparison of inverse magnetic catalysis in the 
two approaches.}


\section{Introduction}
\label{sec:1}
Two of the most important laboratories for the theory of strong interactions exhibit large magnetic fields: firstly, in non-central relativistic 
heavy-ion collisions the magnetic field perpendicular to the collision plane can be as high as $10^{18}\ \mathrm{G}$ \cite{Skokov:2009qp}, and, 
secondly, in certain compact stars called magnetars the surface magnetic field is of the order of $10^{15}\ \mathrm{G}$ \cite{Duncan:1992hi}, 
while the application of the virial theorem suggests that in the interior the magnitude of the magnetic field might reach 
$10^{18}\, \mathrm{G}$ \cite{Lai}. Since this is comparable to the QCD scale $\Lambda_{\mathrm{QCD}}\simeq 200\ \mathrm{MeV}$
[in natural Heaviside-Lorentz units,  $10^{18}\,{\rm G}\, \simeq (140\,{\rm MeV})^2$], the magnetic 
field in these laboratories might have a significant influence on properties governed by the strong interaction. For example, in the case of 
heavy-ion collisions, the magnetic field might be responsible for an observed charge separation, which 
has been attributed to the so-called chiral 
magnetic effect \cite{Kharzeev:2007jp,Fukushima:2008xe,Kharzeev:2009pj}. On the other hand, in compact stars the structural 
composition of the star's interior, i.e., the equation of state, and transport properties could be affected .

Also from a theoretical point of view these two physical systems present a great challenge. Both of them cover regions of the QCD phase diagram that
 are very difficult to study from first principles, 
since the large coupling strength prevents the application of perturbative methods. Relativistic heavy-ion collisions explore 
the phase diagram at low chemical potential and intermediate temperature (of the order of the QCD scale). 
This region is best tackled by lattice 
QCD, which has in the recent years been able to quantify the equilibrium properties in this regime \cite{Aoki:2006br,Aoki:2006we}. 
For transport phenomena, however, lattice simulations are not well suited. Here, 
the application of the AdS/CFT correspondence \cite{Maldacena:1997re}, 
a method developed in string theory, has contributed the celebrated result for the ratio of shear viscosity $\eta$ over entropy density $s$
\cite{Policastro:2001yc}. The result, $\eta/s=1/4\pi$, is currently unrivalled by other methods and appears to agree very well with experimental 
data. Furthermore, this value has been conjectured to be a lower bound for all isotropic fluids \cite{Kovtun:2004de}. (This 
bound has been lowered in higher-derivative gravity duals \cite{Brigante:2007nu}, while in anisotropic fluids there appears to be no lower bound 
\cite{Rebhan:2011vd}.)
Unfortunately, the region of the phase diagram relevant for compact stars, where the temperature is low and the quark chemical potential is of 
the order of the QCD scale, is inaccessible for lattice simulations due to the so-called sign problem. Here one has to rely on extrapolations
(down in density) from perturbative calculations or extrapolations (up in density) from nuclear physics or on suitable models, two of which will be of relevance for this article: the Nambu--Jona-Lasinio (NJL) model 
\cite{Nambu:1961tp,Nambu:1961fr} and the Sakai--Sugimoto model \cite{Sakai:2004cn,Sakai:2005yt}.

In this review we focus on the effect of a background magnetic field on the chiral phase transition of QCD. The Lagrangian of QCD with 
$N_f$ flavors exhibits an approximate global $U(N_f)_L\times U(N_f)_R$ symmetry group, which is broken down to a global 
$SU(N_f)_L\times SU(N_f)_R\times U(1)_{L+R}$ by the axial anomaly. At small temperatures and chemical potentials, i.e., in the hadronic phase, 
the chiral symmetry $SU(N_f)_L\times SU(N_f)_R$ is spontaneously broken to $SU(N_f)_{L+R}$ through the formation of a quark--anti-quark 
condensate. In this scheme the light mesons are understood as the (pseudo-)Goldstone modes corresponding to the broken generators of the symmetry group. By turning on a chemical potential one introduces an asymmetry between quarks and anti-quarks and thus exerts a stress on their pairing. As a consequence, one expects to eventually restore chiral symmetry.\footnote{At asymptotically large chemical 
potentials it is 
known from first principles that chiral symmetry is also broken, however via a different mechanism, namely by the formation of a diquark 
condensate in the color-flavor-locked phase \cite{Alford:1998mk,Alford:2007xm}. Whether the hadronic phase is superseded by normal quark matter 
or by CFL or by some other color-superconducting phase is a matter of debate. Here we shall ignore color superconductivity. For the inclusion
of color superconductivity in the context of the chiral phase transition in a magnetic field see Ref.\ \cite{Fayazbakhsh:2010bh}.}

The two models under consideration realize the implementation and breaking of chiral symmetry quite differently. 
In its original formulation, the NJL model was supposed to explain the mass of nucleons via chiral symmetry breaking. 
With the advent of QCD it was reformulated as a model of quarks \cite{Volkov:1984kq,Hatsuda:1984jm}.
It is a non-renormalizable field theory since it approximates the interaction of quarks by a four-point fermion interaction, 
and therefore the results of the model depend on the regularization scheme and on the UV cut-off that is used. Furthermore, the NJL model in
its standard form lacks confinement. In the chiral limit, the Lagrangian of the NJL model is invariant under the same symmetry group as the QCD Lagrangian with massless quarks -- the \textit{global} $SU(N_c)\times U(N_f)_L\times U(N_f)_R$, where $N_c$ denotes the number of colors. The chiral symmetry is broken explicitly by a bare quark mass, which has to be sufficiently small compared to the momentum cut-off. Spontaneous
breaking of chiral symmetry is then realized very similarly to the BCS theory of superconductivity \cite{Bardeen:1957kj}, which actually served as 
a guideline in the development of the model. Nevertheless, there are important differences between the condensation of Cooper pairs in 
a superconductor and the condensation of fermion--anti-fermion pairs. For instance, the presence of a Fermi surface in the former case 
implies the instability with respect to Cooper pairing for arbitrarily weak attractive interactions, while, as we shall see in the NJL model,
there is a finite 
critical coupling that is needed to form a chiral condensate. We shall also see that the analogy becomes better for chiral condensation 
{\it in a magnetic field}.

Our second model, the Sakai--Sugimoto model, is a top-down string-theoretical approach to a holographic dual of large-$N_c$ QCD. It exploits a non-supersymmetric variation of the original gauge-gravity duality conjectured in \cite{Maldacena:1997re} known as the Witten model \cite{Witten:1998zw}. Sakai and Sugimoto introduced fundamental quarks in the chiral limit by placing a stack of $N_f$ probe $D$-branes for the left-handed and anti-$D$-branes for the right-handed sector into the supergravity limit of the Witten model. According to the holographic dictionary, the local gauge symmetry on these 
``flavor branes'' translates into a global symmetry on the boundary, i.e., into the chiral symmetry of the dual field theory. The question of whether the symmetry is intact or broken amounts to asking whether one can perform gauge transformations on $D$-branes and anti-$D$-branes independently or not, i.e., whether the  $D$-branes connect with the anti-$D$-branes in the bulk. Thus the symmetry 
breaking mechanism in the Sakai--Sugimoto model is of geometrical nature. 

In order to understand what effects a magnetic field might have on the formation of the chiral condensate, let us recapitulate the 
general discussion found in \cite{Gusynin:1994xp}. Calculating the chiral condensate in field theory amounts to calculating a fermion loop. Let the bare fermion mass be finite for the moment and regularize the UV divergence via a cut-off in some suitable scheme, e.g., Schwinger's proper time method. In the presence of a magnetic field one has to take Landau quantization of the transverse momentum of the charged fermions into account. It turns out that if one performs the chiral limit on the result, an IR singularity appears, which can be shown to originate from the lowest Landau level. 
As a consequence, a mass gap is dynamically generated in order to avoid this IR singularity. The precise form of the gap is of course dictated by the form of the interactions in the model under consideration. This effect -- termed magnetic catalysis -- was first found in the Gross-Neveu model \cite{Klimenko:1990rh,Klimenko:1992ch} and later on in several NJL model calculations \cite{Gusynin:1994xp,Gusynin:1994re,Gusynin:1994va,Fukushima:2012xw} and in QED \cite{Gusynin:1995gt} as well as in holographic approaches \cite{Filev:2007gb,Erdmenger:2007bn,Filev:2009xp,Filev:2010pm,Callebaut:2011ab,Bolognesi:2011un,Erdmenger:2011bw}. 
It also plays an important role in the context of graphene \cite{Gusynin:2006gn,Gorbar:2008hu}. 
For QCD, it was found in a lattice calculation (however, with unphysical quark masses) that the critical temperature increases with 
the magnetic field \cite{D'Elia:2010nq}, 
in accordance with magnetic catalysis. However, recently the Budapest-Wuppertal collaboration found (with physical quark masses) that the maximum of the 
quark susceptibility drops 
significantly at temperatures about $140\ \mathrm{MeV}$ under the influence of a magnetic field \cite{Bali:2011qj}, i.e., 
the opposite effect was observed. It remains an open and interesting question what prevents magnetic catalysis to persist for larger temperatures 
in QCD.

This article is mostly, but not exclusively, a review of existing work in the field-theoretical NJL and the holographic Sakai-Sugimoto model. 
In Sec.\ \ref{sec:2}, we review the effect of a magnetic field on chiral symmetry breaking in the NJL model in a pedagogical way, 
starting from the simplest case without magnetic field. This section also contains several new, so far unpublished, aspects, for instance the 
analytic approximations and related discussions regarding inverse magnetic catalysis in Sec.\ \ref{sec:IMCNJL}.
After a pedagogical introduction to the Sakai-Sugimoto model in Sec.\ \ref{sec:introSS}, we discuss its limit of a small asymptotic 
separation of the flavor branes and map out the critical surface of chiral symmetry breaking in the parameter space of temperature, chemical 
potential and magnetic field. 
This part is mostly a review of our own works \cite{Preis:2010cq} and \cite{Preis:2011sp}, with emphasis on the analytic approximations
of the results and their comparison to the field-theoretical analogues.


\section{Chiral phase transition in the Nambu--Jona-Lasinio model}
\label{sec:2}
We start from the standard Lagrangian of the NJL model (for an overview over the various NJL--type models see \cite{Buballa:2003qv}),
\begin{equation}
 \mathcal{L}=\overline{\psi}(\I\gamma^\mu D_\mu-m+\mu\gamma_0)\psi+G\left[\left(\overline{\psi}\psi\right)^2
+\left(\overline{\psi}\gamma_5 \psi\right)^2\right] \, .
\end{equation}
We restrict ourselves to $N_f=1$; $\mu$ denotes the quark chemical potential, $m$ the bare quark mass, 
$D_\mu=\partial_\mu+iqA_\mu$ the covariant derivative with $q$ the electric charge of the quarks and the electromagnetic gauge potential 
$A_\mu=(\phi,-\vec{A})$. 
We shall work with imaginary (Euclidean) time $\tau=-ix^0$ compactified on a circle, the circumference of which is identified with 
the inverse of the temperature $T$. 

As a next step, we assume the pseudoscalar condensate to vanish, $\langle\overline{\psi}\gamma_5\psi\rangle\equiv0$, 
and apply the mean-field approximation,
\begin{eqnarray}
\left(\overline{\psi}\psi\right)^2&\simeq& -\langle\overline{\psi}\psi\rangle^2+2\langle\overline{\psi}\psi\rangle \overline{\psi}\psi
\, .
\end{eqnarray}
We assume the quark--anti-quark condensate $\langle\overline{\psi}\psi\rangle $ to be homogeneous and isotropic. 
For more general ans\"atze see \cite{Tatsumi:2004dx,Nakano:2004cd}, where a dual chiral density wave (a.k.a. chiral spiral) has 
been discussed, and \cite{Nickel:2009wj} for more general inhomogeneous phases. We define the constituent quark mass as
\begin{equation}
M=m-2G\langle\overline{\psi}\psi\rangle.
\end{equation}
In the following we assume stationarity and apply the temporal gauge fixing condition. 
Therefore, the temporal dependence of the  eigenfunctions of the Dirac operator is simply an exponential, $\E^{\I\omega_n\tau}$, 
with the fermionic Matsubara frequencies $\omega_n=(2n+1)\pi T$. Then, the thermodynamic potential becomes 
\begin{equation} \label{Omega00}
\Omega=-\frac{T}{V}\ln Z=\frac{(M-m)^2}{4G}-\frac{T}{V}\Tr\ln \frac{-\I\omega_n+\mu-\epsilon}{T},
\end{equation}
where $Z$ is the partition function. The trace includes summation over the fermionic Matsubara frequencies and some suitable spectral 
decomposition $\epsilon$ of the Dirac Hamiltonian,
\begin{equation}
 H_D=\gamma^0\vec{\ugamma}\cdot(-\I\vec{\nabla}-q\vec{A})+\gamma^0 M \, .
\end{equation}
In analogy to BCS theory, the equation for minimizing the effective potential with respect to $M$ is called gap equation, which reads
\begin{equation} \label{gapgeneral}
\langle\overline{\psi}\psi\rangle=-\frac{T}{V}\Tr \frac{\gamma^0}{\I\omega_n+\mu-\epsilon}.
\end{equation}
In the context of a background magnetic field, we shall also discuss the axial current. Its expectation value is given by 
\begin{equation}\label{axial00}
 \langle j^\mu_5\rangle=-\frac{T}{V}\Tr \frac{\gamma^0\gamma^\mu\gamma_5}{\I\omega_n+\mu-\epsilon}.
\end{equation}


\subsection{Chiral symmetry breaking without external fields}
\label{sec:noB}

Without external fields, the normalized eigenfunctions of the Dirac Hamiltonian are given (in a Weyl basis) by the momentum eigenfunctions
\begin{eqnarray}
\psi_{e,s,k}(\vec{x})&=&
\frac{1}{\sqrt{V}}\E^{\I\vec{k}\cdot \vec{x}}\frac{1}{\sqrt{2\epsilon_k}}\left(\xi^s\sqrt{\epsilon_k-esk},
\xi^se\sqrt{\epsilon_k+esk}\right)^\mathrm{T} \, ,
\end{eqnarray}
where $\xi^s$ are two-vectors defined by the eigenvalue equation $\hat{\vec{k}}\cdot\vec{\sigma}\xi^s=s\xi^s$ with the usual Pauli matrices
$\sigma^i$, and
where $e\epsilon_k=e\sqrt{k^2+M^2}$ with $e=\pm$ is the eigenvalue of the Dirac Hamiltonian, which in turn is 
two-fold degenerate with respect to $s=\pm$. For the diagonal matrix elements of the gamma matrices in this basis we find
\begin{equation}
 \gamma^0_{e,s,k}=e\frac{M}{\epsilon_k} \, , \qquad \left(\gamma^5\gamma^0\hat{\vec{k}}\cdot\vec{\gamma}\right)_{e,s,k}=s.
\end{equation}
{}From the second relation and Eq.\ (\ref{axial00}) we conclude that the axial current vanishes, $\langle j^i_5\rangle=0$. The reason 
is the spin degeneracy in each state. This will no longer be true in the presence
of a magnetic field, as we shall discuss in Sec.\ \ref{sec:IMCNJL}. 
We can compute the thermodynamic potential and, by inserting the first relation into Eq.\ (\ref{gapgeneral}), the gap equation 
in the thermodynamic limit (at vanishing magnetic field $B$),
\begin{eqnarray}
\Omega_{B=0}&&=\frac{(M-m)^2}{4G}-2\sum_{e=\pm}\int\frac{\D^3k}{(2\pi)^3}\left[\frac{\epsilon_k}{2}+T\ln\left(1+\E^{-\frac{\epsilon_k-e\mu}{T}}
\right)\right]\, , \label{OmegaB0} \\[2ex]
M-m&&=4G\int\frac{\D^3k}{(2\pi)^3}\frac{M}{\epsilon_k}\left[1-f(\epsilon_k-\mu)-f(\epsilon_k+\mu)\right] \, , \label{gapeq0}
\end{eqnarray}
where $f(x)\equiv 1/(e^{x/T}+1)$ is the Fermi-Dirac distribution function.
The (vacuum parts of the) momentum integrals are UV divergent and have to be regularized. Since the NJL  model is non-renormalizable, 
all results, e.g., the magnitude of the gap and the order of phase transitions, will depend on the regulator as well as 
on the regularization scheme. We use the proper time regularization scheme \cite{Schwinger:1951nm}. In this procedure, 
the integrand of divergent expressions is recast into so-called proper time integrals,
\begin{equation}
\left(k^2+b^2\right)^{-a}=\frac{1}{\Gamma(a)}\int_0^\infty\D \tau\, \tau^{a-1}\E^{-\tau(k^2+b^2)} \, , 
\end{equation}
and one then performs the momentum integral before the proper time integral. The UV divergence of the momentum integral
reappears at the lower bound of the proper time integral, which therefore has to be regularized. We set the lower bound to $1/\Lambda^2$.

This yields the thermodynamic potential at zero temperature
\begin{eqnarray}\label{freeenergyvac}
16\pi^2 \Omega_{B=T=0} &=&\frac{2\Lambda^2(M-m)^2}{g}+\Lambda^2\left(\Lambda^2-M^2\right)\E^{-M^2/\Lambda^2}
+M^4\Gamma\left(0,\frac{M^2}{\Lambda^2}\right)\nonumber\\[2ex]
&&-\,2\theta(\mu-M)
\left[\frac{\mu k_F}{3}(2\mu^2-5M^2)+M^4\ln\frac{\mu+k_F}{M}\right] \, ,
\end{eqnarray}
where $\Gamma(a,x)$ 
is the incomplete gamma function, and the 
gap equation
\begin{eqnarray}\label{gap00}
M-m&=& Mg\left[\E^{-M^2/\Lambda^2}-\frac{M^2}{\Lambda^2}\Gamma\left(0,\frac{M^2}{\Lambda^2}\right)\right] \nonumber\\[2ex]
&&-\,2Mg\theta(\mu-M)\left(\frac{\mu k_F}{\Lambda^2}-\frac{M^2}{\Lambda^2}\ln\frac{\mu+k_F}{M}\right) \, , 
\end{eqnarray}
where we have defined the  the Fermi momentum $k_F=\sqrt{\mu^2-M^2}$ and the dimensionless coupling constant
\begin{equation}
g \equiv \frac{G\Lambda^2}{2\pi^2} \, .
\end{equation}
For simplicity we shall discuss the chiral limit $m=0$ in the rest of the paper. In this case, $M=0$ is always a solution to the gap equation. 

For $\mu=0$, the gap equation further simplifies since the term $\propto\theta(\mu-M)$ does not contribute. 
Then, after dividing Eq.\ (\ref{gap00}) by 
$M$ and $g$, its right-hand side is always 
smaller than 1. Therefore, a nontrivial solution for $M$ only exists if the dimensionless coupling constant $g$ is larger than 1.
When it exists, this solution is preferred over the trivial solution, as one can verify with the help of the thermodynamic 
potential (\ref{freeenergyvac}).

\begin{SCfigure}[\sidecaptionrelwidth][!h]
\includegraphics[width=0.6\textwidth]{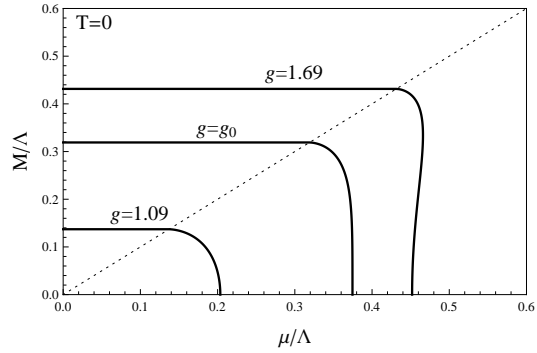}
\caption{The zero-temperature solution to the gap equation for three different values of the coupling $g$. The thin dotted line 
is the line $\mu=M$. The solution becomes multi-valued in the region $\mu>M$ for all couplings larger than $g_0$ with $g_0$ given in 
Eq.\ (\ref{g0}). \vspace{1.5cm}}
\label{gapTBeq0}
\end{SCfigure}

In Fig.\ \ref{gapTBeq0} we show the numerical solution for the gap equation as a function of $\mu$ for three different coupling constants larger 
than 1 (i.e., they all admit a nontrivial solution for $\mu=0$).
For all couplings, there is a certain critical $\mu$ where $M$ goes to zero. By first dividing the 
gap equation by 
$M$ and then setting $M=0$, it is easy to show that this critical $\mu$ is given by 
\begin{equation} \label{mu0}
\frac{\mu_0(g)}{\Lambda}=\frac{1}{\sqrt{2}}\sqrt{1-\frac{1}{g}}  \, .
\end{equation}
If and only if the solution is single-valued, this is the critical $\mu$ at which the (then second-order) phase transition to the chirally 
restored phase occurs.  

Above a certain coupling, the solution becomes multi-valued. The coupling where this qualitative change occurs can be computed as follows.
By differentiating the gap equation with respect to $\mu$ we find
\begin{equation}
\frac{\partial M}{\partial\mu}=-\frac{2k_F}{M\left[\Gamma\left(0,\frac{M^2}{\Lambda^2}\right)-2\ln\frac{\mu+k_F}{M}\right]}. 
\end{equation}
In accordance to the numerical plot, this derivative is infinite for $M=0$. For all couplings for which the solution is multi-valued, there
is another point where the derivative is infinite, which is given by the second pole of the denominator,
\begin{equation}
\mu=M\cosh\frac{\Gamma(0,M^2/\Lambda^2)}{2} \, .
\end{equation}
We can now ask for the value of $g$ at which this point coincides with $\mu_0(g)$ for $M\to 0$. The resulting equation then 
yields the coupling where the multi-valuedness sets in. We find 
\begin{equation}\label{g0}
g_0=\frac{1}{1-\frac{e^{-\gamma_E}}{2}} \simeq 1.390 \, ,
\end{equation}
where $\gamma_E$ is the Euler-Mascheroni constant. In the regime $1<g<g_0$ the chiral phase transition is second order and takes 
place at $\mu_0(g)$. 

\begin{figure}[!h]
\begin{center}
\includegraphics[scale=0.8]{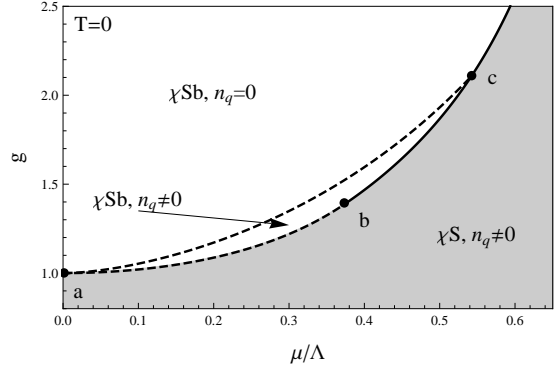}
\end{center}
\caption{The phase diagram at $T=0$ in the $\mu$-$g$-plane. Dashed lines indicate second-order, solid lines
first-order phase transitions. In the shaded region chiral symmetry is restored ($\chi$S). 
The points a, b and c correspond to $(\mu/\Lambda,g)=(0,1)$, $(e^{-\gamma_E/2}/2,g_0)$,  and $(0.542,2.106)$, respectively, with $g_0$ given 
in Eq.\ (\ref{g0}). Between points a and b the transition line is given by $\mu_0(g)$ from Eq.\ (\ref{mu0}). The dashed line between a and c
indicates the onset of a finite quark number density $n_q$ within the chirally broken phase ($\chi$Sb).}
\label{TBeq0pd}
\end{figure}

For couplings larger than $g_0$ the transition is first order and has to be determined numerically. 
It turns out that the branch with a positive slope is always energetically disfavored. Therefore, in terms of Fig.\ \ref{gapTBeq0}, 
the preferred solution follows the horizontal line $M(\mu=0)$ and, for all multi-valued cases, jumps to zero at a certain chemical potential. 
Whether (and how far) the preferred solution follows the curve into the region $\mu>M$ depends on the coupling. We find numerically that for 
couplings below (above) $g\simeq 2.106$ it does (doesn't). This is a first example of the nontrivial effect of $\mu$ on the 
preferred phase: it is not always the phase with the largest dynamical mass that is favored. 
In more physical terms, for couplings above $g\simeq 2.106$ the chirally broken phase with 
vanishing quark density is directly superseded by the quark matter phase, while for smaller couplings there is a region of finite density
between these two phases. Since for $g>2.106$ there are no complicated effects of the quark density, we can write down a very simple 
expression for the free energy difference between the broken phase and the restored phase, evaluated at the solution of the gap equation 
(and using $M\ll\Lambda$),
\begin{equation}
\Delta\Omega=-\frac{M_0^2\Lambda^2}{16\pi^2}\left(1-\frac{1}{g}\right)+\frac{\mu^4}{12\pi^2} \, ,
\label{LLLfreeenergy}
\end{equation}
with $M_0$ being the (non-analytical) solution to the gap equation for $\mu=0$. This result is very intuitive: the first, negative, term 
is the condensation energy, i.e., the energy gain from the chiral condensate, while the second, positive, term corresponds to the energy 
costs for pairing which must be paid because the chemical potential has separated fermions from anti--fermions. When the costs exceed the gain, 
chiral
symmetry is restored. This determines the phase transition line. Below we shall derive the analogue of this strong-coupling free energy difference 
in the presence of a background magnetic field, see Eq.\ (\ref{imc}). We summarize our discussion of the chiral phase 
transition at $B=T=0$ in Fig.\ \ref{TBeq0pd}.

\begin{figure}[!t]
\begin{minipage}{\textwidth}
\includegraphics[height=3.7cm]{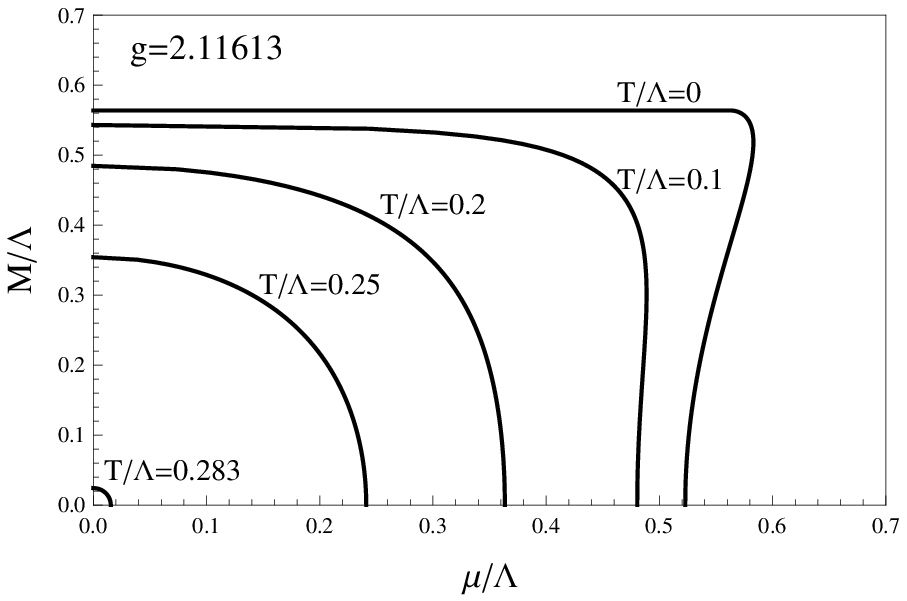}
\hspace{\fill}
\includegraphics[height=3.7cm]{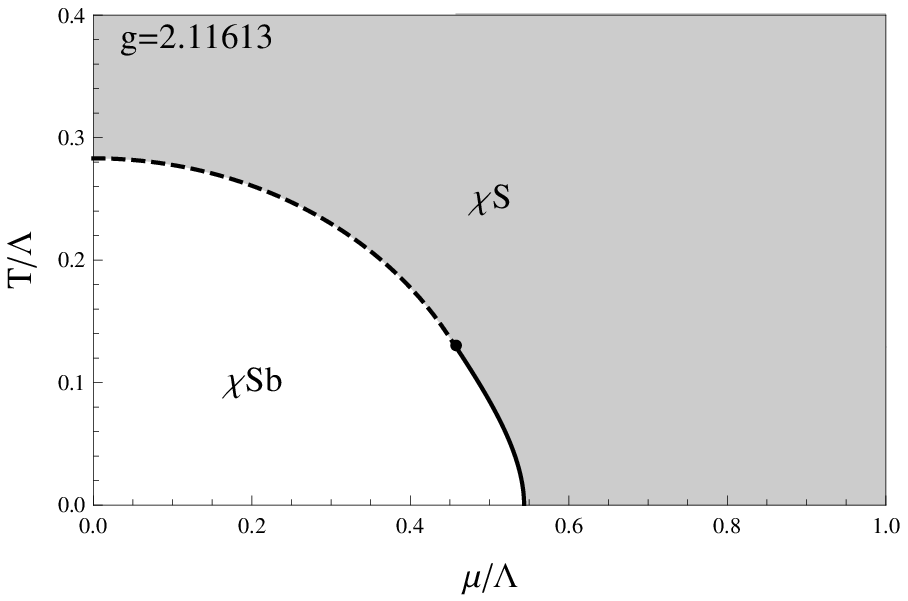}
\end{minipage}
\caption{Finite-temperature effects on the chiral phase transition in the NJL model. 
Left panel: the gap as a function of the chemical potential for a given coupling strength and different values of temperature. 
Right panel: the phase diagram in the $\mu$-$T$-plane for the same coupling. The (dashed) second-order phase transition line is given by the 
analytic expression (\ref{Tc}). }
\label{finiteT}
\end{figure}

For nonzero temperatures, we need to solve the gap equation (\ref{gapeq0}) [with the regularization of the vacuum part shown 
in Eq.\ (\ref{gap00})] numerically. The result for various temperatures and a large coupling (larger than that of point c in Fig.\ \ref{TBeq0pd}) 
is shown in the left panel of Fig.\ \ref{finiteT}.
In general, the temperature decreases the gap. Moreover, the temperature can also change the order of the chiral phase 
transition by removing the multi-valuedness of the solution to the gap equation. The critical temperature of the chiral phase transition 
in the $T$-$\mu$ phase diagram is shown in the right panel of Fig.~\ref{finiteT}.
The critical point moves towards higher temperatures with increasing coupling. If the phase transition is second order it is possible to find a 
closed form for the critical temperature. To this end, one divides Eq.\ (\ref{gapeq0}) (with $m=0$) by $M$ and then sets $M=0$ in the remaining 
equation. Then, solving for $T$ yields the critical temperature 
\begin{equation} \label{Tc}
\frac{T_c(\mu)}{\Lambda}=\sqrt{\frac{3}{2\pi^2}}\sqrt{1-\frac{1}{g}-2\frac{\mu^2}{\Lambda^2}} \, . 
\end{equation}

\subsection{Chiral symmetry breaking in the presence of a magnetic field}

\subsubsection{Structure of the fermion states in a background magnetic field}

Let us consider a homogeneous background magnetic field $\vec{B}=(0,0,B)$ by choosing the Landau gauge fixing condition with $\vec{A}=(-yB,0,0)$. 
Within this ansatz, the eigenfunctions of the Hamiltonian are proportional to $\exp[\I(\omega_n\tau+k_xx+k_zz)]$.
Using this, we split the Dirac Hamiltonian in a longitudinal and a transverse part with respect to the direction of the magnetic field, 
$H_\mathrm{D}=H_\mathrm{L}+H_\mathrm{T}$, where
\begin{eqnarray}
H_\mathrm{L}&&=\gamma^0\gamma^3 k_z+\gamma^0M\, , \qquad
H_\mathrm{T}=\sgn (q) 
\sqrt{2\left|q\right|B}\left(\begin{array}{cc} -1 & 0 \\ \ \ 0 & 1\end{array}\right)\otimes\left(\begin{array}{cc} 0 &\ a^\dagger 
\\ a & 0 \end{array}\right)\, ,
\end{eqnarray}
with  
\begin{eqnarray}
a&& \equiv \sqrt{\frac{\left|q\right|B}{2}}\, \xi
+\sgn(q)\I\frac{1}{\sqrt{2\left|q\right|B}}\left(-\I\partial_\xi\right)\,, \qquad \xi\equiv y+\frac{k_x}{qB} \, .
\end{eqnarray}
We see that $a$ is the annihilation (creation for $q<0$) operator of the quantum mechanical oscillator, which gives rise to the Landau quantization 
of the energy spectrum of a charged fermion moving in a background magnetic field.
For $q>0$, the orthogonalized eigenfunctions of the Hamiltonian are given by
\begin{eqnarray}
\psi_{k_x,k_z,\ell}^{e,s}(\vec{x})&&=\frac{\E^{\I(k_zz+k_xx)}}{\sqrt{L_{x}L_{z}}}\frac{1}{2\sqrt{\kappa_{k_z,\ell}\epsilon_{k_z,\ell}}}
\left(\begin{array}{c} 
s\sqrt{\kappa_{k_z,\ell}+sk_z}\sqrt{\epsilon_{k_z,\ell}-s\kappa_{k_z,\ell}}  \,\langle \xi\vert\ell\rangle \\[1ex] 
\sqrt{\kappa_{k_z,\ell}-sk_z}\sqrt{\epsilon_{k_z,\ell}-s\kappa_{k_z,\ell}}   \,\langle \xi\vert\ell-1\rangle \\[1ex] 
es\sqrt{\kappa_{k_z,\ell}+sk_z}\sqrt{\epsilon_{k_z,\ell}+s\kappa_{k_z,\ell}} \,\langle \xi\vert\ell\rangle \\[1ex] 
e\sqrt{\kappa_{k_z,\ell}-sk_z}\sqrt{\epsilon_{k_z,\ell}+s\kappa_{k_z,\ell}}  \,\langle \xi\vert\ell-1\rangle 
\end{array}\right) \quad\;\; \label{psiplus}
\end{eqnarray}
where $\ell=0,1,2,3, \ldots$ denotes the Landau level, where
\begin{eqnarray}
 \langle \xi\vert\ell\rangle&&=\frac{1}{\sqrt{2^\ell\ell!}}\left(\frac{\left|q\right|B}{\pi}\right)^{1/4}\E^{-\left|q\right|B\xi^2/2}H_\ell
\left(\sqrt{\left|q\right|B}\xi\right)\, ,
\end{eqnarray}
$\langle \xi\vert -1 \rangle\equiv0$, and
\begin{eqnarray}
\epsilon_{k_z,\ell}&&= \sqrt{k_z^2+M^2+2\left|q\right|B\ell} \,, \qquad \kappa_{k_z,\ell}=\sqrt{k_z^2+2|q|B\ell}.
\end{eqnarray}
Here, $H_\ell$ is the $\ell$th Hermite polynomial and $L_{i}$ the length of a 
box with volume $V$ in the $i$th direction. In order to obtain the eigenfunctions for the case $q<0$, one simply replaces 
$\langle \xi\vert\ell\rangle$ with $\langle \xi\vert\ell-1\rangle$ and vice versa. For the diagonal matrix elements of $\gamma^0$ and 
$\gamma^0\gamma^3\gamma_5$ we find
\begin{eqnarray}
\gamma^0_{e,s,k_z,\ell}=e\frac{M}{\epsilon_{k_z,\ell}} \, , \qquad  (\gamma^0\gamma^3\gamma_5)_{e,s,k_z,\ell}=\sgn(q)
\frac{sk_z}{\kappa_{k_z,\ell}} \, . \label{axialcurrent1}
\end{eqnarray}
{}From Eq.\ (\ref{psiplus}) we see that in the lowest Landau level (LLL) $\ell=0$ only the $\sgn(q)\ s=1$-states survive, which are also 
eigenstates of the spin operator $\Sigma_3=\gamma^0\gamma^3\gamma_5$ as well as zero-eigenmodes of $H_\mathrm{T}$. Therefore, the dynamics of the 
LLL becomes effectively $1+1$-dimensional. Moreover, in the limit $M\rightarrow 0$ for $\sgn(q) e\ k_z>0\ (<0)$ these states are right- (left-) 
handed only. This is an indication that the magnetic field induces an axial current \cite{Metlitski:2005pr}. More precisely, due to the sum over $s$
in the axial current (\ref{axial00}), the relation (\ref{axialcurrent1}) shows that only the LLL level contributes. Due to the sum over 
$e$ there can only be a finite contribution if $\mu\neq 0$. Since we have put the fermions into a box with volume $V=L_{x}L_{y}L_{z}$, the range of $y$ is restricted to $[-L_{y}/2,L_{y}/2]$ and therefore $k_{x,{\mathrm{max}}}-k_{x,{\mathrm{min}}}=L_{y}\left|q\right|B$ since we have absorbed $k_x$ into the new coordinate $\xi$. Hence, because of $\Delta k_x=2\pi/L_{x}$, each energy level for given $e$, $k_z$, $s$ and $\ell$ has a 
degeneracy of $L_{x}L_{y}\left|q\right|B/(2\pi)$. In two cases the result for the axial current along the magnetic field can be given in closed form,
\begin{eqnarray}
 M=0, \forall\ T:&&\ \ \langle j^3_5\rangle=\frac{qB\mu}{2\pi^2} \, , \label{NJLaxialcurrent}\\
 T=0, \forall\ M<\mu:&&\ \ \langle j^3_5\rangle=\frac{qB\sqrt{\mu^2-M^2}}{2\pi^2} \, .
\end{eqnarray}
The prefactor $\left|q\right|B/(2\pi)$ found by phase space considerations has a very special role here. It is the difference of 
the number of zero-eigenmodes of $H_\mathrm{T}$ with $s=1$ and $s=-1$ respectively. This is a topological result since it is given by the index of each $2\times 2$ block of $H_\mathrm{T}$, which in turn is linked to the Euclidean chiral anomaly in two dimensions via the index theorem. Furthermore, the first result is independent of $T$ which is a special feature of massless $1+1$ dimensional fermions and hence reflects the effective dimensional reduction.

\subsubsection{Magnetic Catalysis}
\label{sec:MC}

Let us return to chiral symmetry breaking, now in the presence of a magnetic field. 
The thermodynamic potential and the gap equation read
\begin{eqnarray}
\Omega&&=\frac{M^2}{4G}-\frac{\left|q\right|B}{2\pi}\sum_{e=\pm}\sum_{\ell=0}^\infty\alpha_\ell\int_{-\infty}^\infty \frac{\D k_z}{2\pi}
\left[\frac{\epsilon_{k_z,\ell}}{2}+T\ln\left(1+\E^{-\frac{\epsilon_{k_z,\ell}-e\mu}{T}}\right)\right] \, , \label{OmTB}\\[2ex]
M&&=2G\frac{\left|q\right|B}{2\pi}\sum_{\ell=0}^\infty\alpha_\ell\int_{-\infty}^\infty
\frac{\D k_z}{2\pi}\frac{M}{\epsilon_{k_z,\ell}}\left[1-f(\epsilon_{k_z,\ell}-\mu)-f(\epsilon_{k_z,\ell}+\mu)\right] \, ,
\label{gapeqBvac}
\end{eqnarray}
where $\alpha_\ell\equiv 2-\delta_{0\ell}$. Comparing with the corresponding $B=0$ expressions in Eqs.\ (\ref{OmegaB0}) and (\ref{gapeq0}), we see 
that the effect of the magnetic field is to replace $\epsilon_k\to \epsilon_{k_z,\ell}$ and
\begin{eqnarray}
2\int\frac{\D^3k}{(2\pi)^3} \to \frac{\left|q\right|B}{2\pi}\sum_{\ell=0}^\infty\alpha_\ell\int_{-\infty}^\infty \frac{\D k_z}{2\pi} \, .
\end{eqnarray}
Using again proper time regularization, the thermodynamic potential at vanishing temperature  becomes
\begin{eqnarray} \label{OmegaT0}
\Omega_{T=0} &=& \Omega_{\mu=T=B=0} - \frac{(qB)^2}{2\pi^2}\left[\frac{x^4}{4}(3-2\ln x)+\frac{x}{2}\left(\ln\frac{x}{2\pi}-1\right)+\psi^{(-2)}(x)
\right] \nonumber\\[2ex] 
&& -\frac{|q|B}{4\pi^2}\theta(\mu-M)\sum_{\ell=0}^{\ell_{\mathrm{max}}}\alpha_\ell
\left(\mu k_{F,\ell}-M_\ell^2\ln\frac{\mu+k_{F,\ell}}{M_\ell}\right)
\end{eqnarray}
Here, $\Omega_{\mu=T=B=0}$ is the vacuum part from Eq.\ (\ref{freeenergyvac}), $\psi^{(n)}$ the $n$-th polygamma function (analytically
continued to negative values of $n$), we have abbreviated $x\equiv M^2/(2|q|B)$, and 
\begin{eqnarray}
 M_\ell\equiv \sqrt{M^2+2|q|B\ell} \,  , \;\;\;\; k_{F,\ell}\equiv\sqrt{\mu^2-M_\ell^2} \, , \;\;\;\; \ell_{\rm max} \equiv 
\left\lfloor\frac{\mu^2-M^2}{2|q|B}\right
\rfloor.
\end{eqnarray}
Different regularization schemes -- compare for instance with \cite{Menezes:2008qt}, where dimensional regularization is used -- only differ in the 
$B=0$ result and in (divergent) terms that depend on $B$ but are constant in $M$, which are omitted. The latter can be viewed as renormalizing the 
energy content coming solely from the magnetic field.

The corresponding gap equation is
\begin{eqnarray} \label{gapT0}
M&=&Mg\left[\E^{-M^2/\Lambda^2}-\frac{M^2}{\Lambda^2}\Gamma\left(0,\frac{M^2}{\Lambda^2}\right)\right] \nonumber\\[2ex]
&& +\, 2Mg \frac{|q|B}{\Lambda^2}\left[\left(\frac{1}{2}-x\right)\ln x + x +\ln\Gamma(x)-\frac{1}{2}\ln 2\pi \right]  \nonumber\\[2ex]
&& -2Mg\frac{|q|B}{\Lambda^2}\sum_{\ell=0}^{\ell_{\mathrm{max}}}\alpha_\ell\ln\frac{\mu+k_{F,\ell}}{M_\ell}\theta(\mu-M)\, .
\end{eqnarray}
Let us first consider the case $\mu=0$, i.e., we can ignore the terms $\propto \theta(\mu-M)$ in 
Eqs.\ (\ref{OmegaT0}) and (\ref{gapT0}). 
For small coupling $g\ll 1$, the dynamical mass squared will be much smaller than the magnetic field, $M^2\ll |q|B$. Then, with $M\ll \Lambda$, 
the gap equation becomes  
\begin{equation}
\frac{1}{g}\simeq \frac{2|q|B}{\Lambda^2}\ln\sqrt{\frac{|q|B}{\pi M^2}}.\label{gapequsmallg}
\end{equation}
Now, there is a nontrivial solution for arbitrarily small $g$. This is in contrast to the case $B=0$ where chiral symmetry can be broken 
only for $g>1$. The solution is obviously
\begin{equation} \label{gapsmallg}
M\simeq \sqrt{\frac{|q|B}{\pi}}\E^{-\frac{\pi^2}{|q|BG}}.
\end{equation}
This qualitative effect of the magnetic field on chiral symmetry breaking was termed ``magnetic catalysis'' in \cite{Gusynin:1994re}
and was since observed in numerous different models. Interestingly, as already mentioned in the introduction, this effect stems mainly 
from the physics in the LLL. 
In order to show that, one omits all contributions from $\ell>0$ in (\ref{gapeqBvac}) and cuts off the momentum integral at 
$\sqrt{|q|B/4\pi}$, since below that cut-off the LLL dominates. Then, one obtains exactly the result (\ref{gapsmallg}).
Furthermore, the logarithmic IR singularity in (\ref{gapequsmallg}) regulated by the dynamically generated mass is precisely due to the LLL 
contribution and its $1+1$ dimensional nature. The form of the gap in the weak coupling limit is reminiscent of the BCS gap in a superconductor 
\cite{Bardeen:1957mv}. In both expressions for the respective gap the relevant density of states appears in the denominator of the exponent. Here it is the density of states of the massless fermions at $\epsilon_{k_z,\ell=0}=0$, whereas in the BCS gap it is the 
density of states at the Fermi surface. In both cases the dynamics is essentially $1+1$-dimensional. While in BCS theory
this effective dimensional reduction is a consequence of the Fermi surface, here it is provided by 
the magnetic field. Note that the dimensional reduction 
is not in conflict with the Mermin--Wagner--Coleman theorem that states that no spontaneous symmetry breaking can occur in $1+1$ dimensions. 
The reason is that the Nambu--Goldstone modes are neutral, and hence their motion is not restricted by the magnetic field. 
At extremely large magnetic fields the internal structure of these modes can be resolved which might 
invalidate this argument \cite{Fukushimatalk}.

\begin{figure}[!t]
\begin{minipage}{\textwidth}
\includegraphics[height=3.63cm]{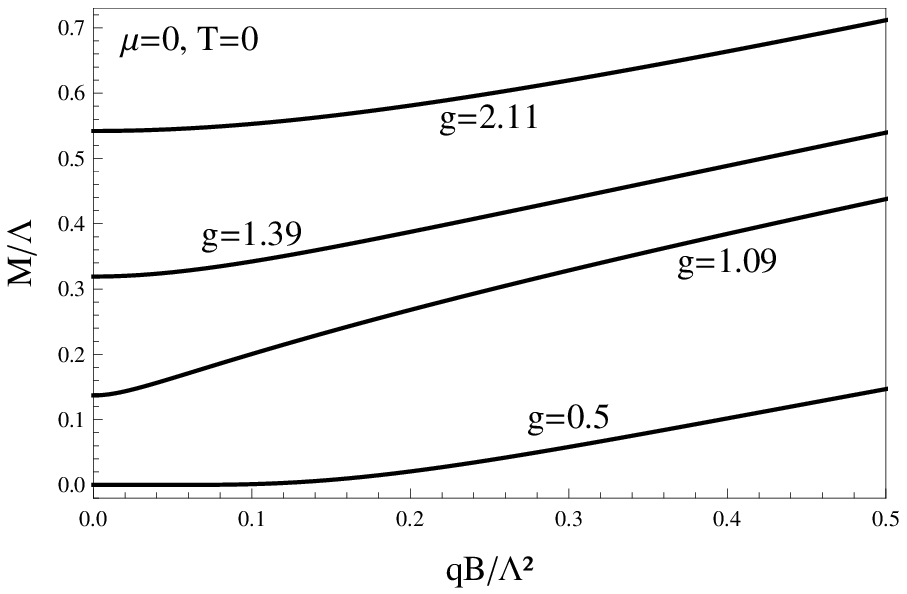}
\hspace{\fill}
\includegraphics[height=3.7cm]{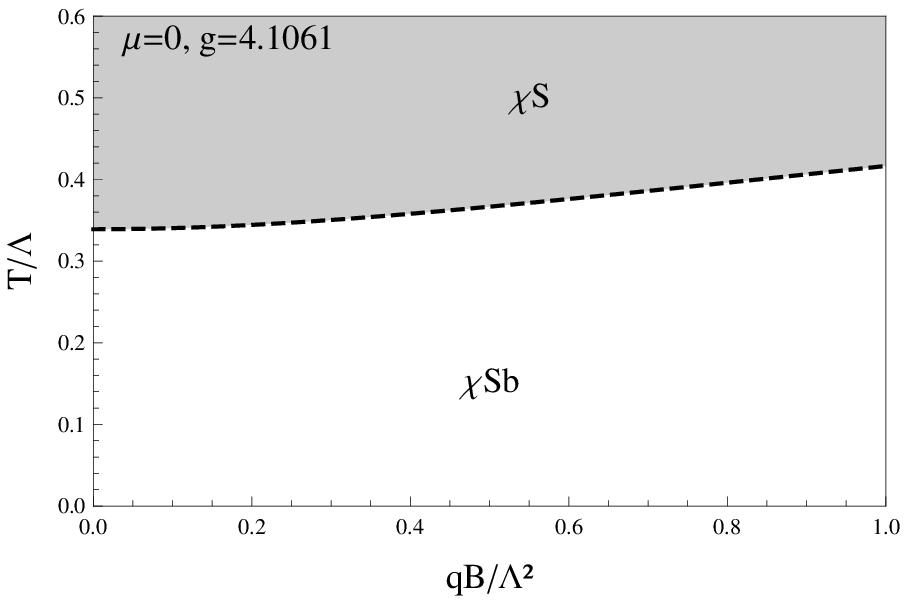}
\end{minipage}
\caption{Effects of magnetic catalysis on the dynamical mass $M$ and the critical temperature. Left: the gap at $T=\mu=0$ for 
different couplings. The lowest coupling shown corresponds to a subcritical coupling at $B=0$, i.e., its nonzero value is solely induced by $B$. 
Its behavior at small $B$ is given by the exponential in Eq.\ (\ref{gapsmallg}).
Right: the critical temperature for chiral symmetry restoration as a function of $B$.}
\label{gapTmueq0}
\end{figure}

We show the numerical solution of the gap equation for various coupling strengths for $T=\mu=0$ in the left panel of Fig.\  \ref{gapTmueq0}.
Magnetic catalysis also manifests itself in the critical temperature for chiral symmetry restoration, which, at $\mu=0$,  monotonically increases 
with increasing magnetic field, see right panel of Fig.\ \ref{gapTmueq0}. 

\subsubsection{Inverse magnetic catalysis}
\label{sec:IMCNJL}

We now include the contributions from a nonvanishing chemical potential $\mu$. 
First we discuss the case of weak coupling which corresponds to $M^2 \ll |q|B$. Since the chiral phase transition can be expected to 
occur at chemical potentials of the order of the mass gap, we may thus also assume $\mu^2\ll|q|B$ (we are not interested in the physics far 
beyond the phase transition). As a consequence, we can employ the lowest Landau level approximation, i.e., drop the contribution of all higher 
Landau levels. Then, from Eq.\ (\ref{OmegaT0}) we conclude that the difference of the thermodynamical potentials of the chirally broken phase 
and the quark matter phase is
\begin{eqnarray}\label{Clogston}
\Delta\Omega&\simeq&\frac{|q|B}{4\pi^2}\left(\mu^2-\frac{M^2}{2}\right)-\frac{|q|B}{4\pi^2}\mu k_{F,0}\theta(\mu-M) \nonumber\\[2ex]
&&+ \frac{\Lambda^2M^2}{8\pi^2}\underbrace{\left(\frac{1}{g}-\frac{2|q|B}{\Lambda^2}\ln \sqrt{\frac{|q|B}{\pi M^2}}
+\frac{2|q|B}{\Lambda^2}\theta(\mu-M)\ln\frac{\mu+k_{F,0}}{M}\right)}_{=0\; {\rm via}\; {\rm gap}\;  {\rm equation}} \, .
\end{eqnarray}
Again, we find a very interesting analogy to superconductivity: the resulting expression is exactly the same as for a BCS superconductor with 
mismatched Fermi momenta -- first discussed by Clogston \cite{Clogston:1962zz} and Chandrasekhar \cite{chandrasekhar} --  
after $M$ is replaced with the superconducting gap $\Delta$, $|q|B$ with the 
average Fermi momentum (squared) of the constituents of a Cooper pair, and $\mu$ with the difference of the respective Fermi momenta. 
(Note that again the degeneracy factor of the LLL emulates the role of the Fermi surface.) 

To discuss the meaning of $\Delta\Omega$ for the chiral phase transition, let us first consider the case of a fixed magnetic field $B$ and
start from $\mu=0$, i.e., in the chirally broken phase. Upon increasing $\mu$, we will reach the point $\mu=M/\sqrt{2}$ where $\Delta\Omega$
changes its sign and thus the phase transition to the chirally restored phase occurs. This point is, in the context of superconductivity, called the
Clogston limit. It occurs before the second term has a chance to 
contribute since still $\mu<M$. Now, more importantly for our purpose, let us again start in the chirally broken phase, i.e., from 
$\Delta\Omega<0$, but now we increase the magnetic field at fixed $\mu$ (as we have just seen, for the discussion of the phase transition
we may assume $\mu<M$ and thus ignore the term $\propto\theta(\mu-M)$). 
Since we have started from a negative $\mu^2-M^2/2$, increasing the magnetic field can only make $\Delta\Omega$ more negative because the dynamical 
mass increases with $B$. Consequently, the magnetic field only brings us ``deeper'' into the chirally broken phase. This is what we have 
expected from magnetic catalysis. 

\begin{figure}[!h]
\begin{center}
\includegraphics[scale=0.9]{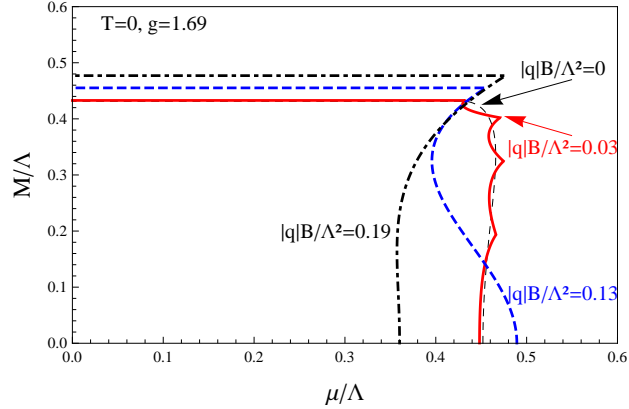}
\end{center}
\caption{The zero-temperature dynamical mass as a function of the chemical potential for different values of the magnetic field. 
For the lowest nonzero value of $|q|B$ shown (solid line), Landau
level oscillations can be seen. The magnetic field for the two other curves (dashed and dashed-dotted lines) is sufficiently large to suppress
all Landau levels except for the lowest.} 
\label{gapTeq0}
\end{figure}

However, as we will now explain, for $g>1$ and finite chemical potential this expectation is incorrect. We shall rather find that, for 
intermediate values of the magnetic field, an increasing magnetic field {\it does} restore chiral symmetry. Let us, to this
end, first discuss the numerical solution of the gap equation, see Fig.\ \ref{gapTeq0}. 
Due to the sum over the Landau levels, the gap exhibits the well-known de Haas--van Alphen oscillations. Similar to the behavior found for 
$B=0$, only the branches with a negative slope of $M(\mu)$ can be energetically preferred. Depending on the specific value of $g$ there 
might be several phase transitions within the gapped phase into regions with $\mu>M$, i.e., with a finite quark number density, before 
entering the restored phase $M=0$. In general it is also possible that the order of the phase transition into the restored phase 
oscillates
between first and second order upon varying $B$: in the example shown in the plot, at vanishing magnetic field the phase transition is 
first order, while at $|q|B/\Lambda^2=0.13$ it is second order and at $|q|B/\Lambda^2=0.19$ again first order. 
We also see that the dashed (blue) curve for the {\it lower} magnetic field reaches farther in the $\mu$ direction than the 
dashed-dotted (black) curve for the {\it larger} magnetic field. This is a surprise from the point of view of magnetic catalysis: it 
seems to indicate that the critical chemical potential for chiral symmetry breaking can decrease with increasing magnetic field. We discuss 
this ``inverse magnetic catalysis'' in more detail now.

\begin{figure}[!t]
\begin{center}
\includegraphics[scale=0.9]{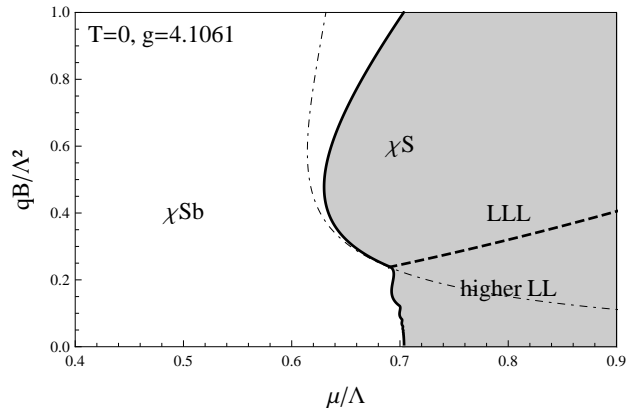}
\end{center}
\caption{Zero-temperature chiral phase transition in the plane of magnetic field and quark chemical potential at a rather large value
of the coupling constant such that the phase transition is first order for all magnetic fields. (For smaller values the shape of the transition 
line is similar, but the order can vary between first and second.) Apart from oscillations at small $B$ due to higher Landau levels in the 
chirally restored phase, the critical chemical potential {\it decreases} up to $qB/\Lambda^2 \simeq 0.5$, see explanation in the text.
The dashed-dotted line is the approximation to the phase transition line from Eq.\ (\ref{imc}). }
\label{Teq0pd}
\end{figure}

To this end, let us consider the ``cleaner" case of sufficiently 
large couplings where symmetry restoration happens in the region $\mu<M$ for all magnetic fields. In this case, oscillations of the critical 
line in the phase diagram originate solely from the restored phase (not from the solution of the gap equation), and the phase transition 
is always first order. The numerically obtained phase diagram for such a case is shown in Fig.\ \ref{Teq0pd}. From the arguments
in the previous subsection, 
one might have expected that magnetic catalysis leads to a monotonically increasing critical chemical potential as a function of $B$ (just like
the critical temperature in the right panel of Fig.\ \ref{gapTmueq0}). However, this is not the case: there is a region in the phase diagram where, 
upon increasing $B$ at fixed $\mu$, chiral symmetry is {\it restored}, in contrast to the weak-coupling case discussed below Eq.\ (\ref{Clogston}).

In order to understand this phenomenon, let us derive an analytic expression for $\Delta\Omega$, analogous to the weak-coupling case. 
As discussed, 
for the given large coupling, the solution to the gap equation is simply given by the $\mu=0$ solution. For small magnetic fields, $|q|B\ll M^2$, 
we can expand the solution up to second order in the magnetic field,
\begin{eqnarray}
M\simeq M_0 \left[1+\frac{(qB)^2}{6M_0^4\Gamma(0,M_0^2/\Lambda^2)}\right] \, ,
\end{eqnarray}
where $M_0$ is the solution for $\mu=B=0$. Inserting this solution into Eq.\ (\ref{OmegaT0}), we obtain the free energy for the chirally broken 
phase up to second order in $B$. The free energy for the chirally restored phase is, although we can set $M=0$, complicated due to the sum
over Landau levels. Let us therefore ignore the higher Landau levels. This seems to contradict our assumption 
of a small magnetic field which we have made for the chirally broken phase. Nevertheless, we shall see that the phase transition line obtained 
from this approximation reproduces the full numerical line in a region of intermediate magnetic fields. Since this is exactly the region 
where the ``back bending'' of the phase transition line is most pronounced, this serves our purpose to capture the main physics of 
the inverse magnetic catalysis. With $M_0\ll\Lambda$, the resulting free energy difference is
\begin{eqnarray}\label{imc}
\Delta\Omega &\simeq& -\frac{M_0^2\Lambda^2}{16\pi^2}\left(1-\frac{1}{g}\right)
+\frac{|q|B}{4\pi^2}\mu^2-\frac{(qB)^2}{24\pi^2}\left[1-12\zeta'(-1) +\ln\frac{M_0^2}{2|q|B}\right]  \, .
\end{eqnarray}
(This is the generalization of Eq.\ (\ref{LLLfreeenergy}) to nonzero (but small) magnetic fields.)
This expression allows for a qualitatively different phase transition line compared to the weak-coupling limit (\ref{Clogston}) for the 
following reason. The term linear in $B$ corresponds to the cost in free energy to form a fermion--anti-fermion condensate 
at finite $\mu$. 
Importantly, this cost depends not only on $\mu$, but also on the magnetic field. This is also true at weak coupling. However, in that case, the gain from the condensation 
energy was also linear in $B$. This is different here: now, if we start from the chirally broken phase, i.e., from $\Delta\Omega<0$, increasing
the magnetic field {\it can} lead to a change of sign for $\Delta\Omega$ and thus restore chiral symmetry. This is what we have
termed inverse magnetic catalysis in \cite{Preis:2010cq}. In this reference, we have also explained that the physical picture can be
understood once again in analogy to superconductivity, where, in the presence of a mismatch in Fermi momenta, it is useful to think of a 
fictitious state where both fermion species are filled up to a common Fermi momentum. Creating such a state costs free energy which may or may not
be compensated by condensation. The point of inverse magnetic catalysis is that creating such a fictitious state (where fermions
and anti-fermions are not separated by $\mu$) becomes more costly with increasing $B$, while $B$ still enhances the dynamical gap due
to magnetic catalysis. The magnetic field thus plays an ambivalent role by counteracting its own catalysing 
effect.

This effect was first observed in the NJL model in \cite{Ebert:1999ht} at $T=0$ and in \cite{Inagaki:2003yi} for 
the full three dimensional $T$-$\mu$-$B$ parameter space, and has been confirmed in various other calculations  
\cite{Fayazbakhsh:2010bh,Boomsma:2009yk,Chatterjee:2011ry,Avancini:2012ee,Andersen:2012bq,Fayazbakhsh:2012vr,Ferrari:2012yw}. 
Only for sufficiently strong magnetic fields the system enters the regime where 
magnetic catalysis is dominant. Typical fits of the model-parameters yield a cut-off of the order of a few hundred $\mathrm{MeV}$ 
\cite{Buballa:2003qv}. Translating this into a scale for the magnetic field shows that the regime of magnetic catalysis is 
beyond the magnetic field strength expected in compact stars, and thus, if there is any observable effect of the magnetic field for the 
phase transition between hadronic and quark matter, it is inverse magnetic catalysis.

\section{Chiral phase transition in the Sakai--Sugimoto model}
\label{sec:3}

\subsection{Introducing the model}
\label{sec:introSS}

The model discussed in this section is based on the conjecture that particular strongly coupled quantum gauge theories are equivalent to 
certain classical gravitational theories in higher dimensions. In the context of string theory, the first realization of this holographic principle known as AdS/CFT correspondence was proposed by Maldacena \cite{Maldacena:1997re}. In a nutshell, it utilizes two different limits of describing so-called D-branes, 
which are dynamical 
objects in string theory that impose Dirichlet boundary conditions on the endpoints of open strings. On the one hand, a stack of $N_c$ 
D-branes hosts a maximally supersymmetric $U(N_c)$ gauge theory coming from the massless excitations of open superstrings; on the other hand, the stack of 
D-branes is a massive object that curves space-time by coupling to gravitons -- coming from the closed strings -- with the strength 
$\lambda\propto g_sN_c$, where $g_s$ denotes the string coupling. Now, let $N_c\rightarrow\infty$ and keep $\lambda$ fixed. 
In the limit $\lambda\ll1$, gravity decouples from the open strings, whose low-energy effective theory is given by the mentioned $U(N_c)$ super Yang--Mills 
theory. In the case of D3-branes, this gauge theory is four-dimensional. In the opposite limit, $\lambda\gg1$, the stack of D-branes 
back-reacts strongly on the background. Gravity far in the asymptotic region also decouples from the system due to the gravitational red shift. 
Therefore, one can zoom in to the near-horizon region of the space-time, which in the case of D3-branes is given by $AdS_5\times S^5$. 
The idea behind the AdS/CFT duality is that the classical 
(super-)gravitational description is fully equivalent to the quantum
theory of the large-$N_c$, large $\lambda$ limit of the 
super-Yang-Mills theory. This particular gauge/gravity duality, which
has passed many nontrivial tests, has since been greatly generalized
and also been used in the form of phenomenological (bottom-up) models.

The Sakai-Sugimoto model \cite{Sakai:2004cn,Sakai:2005yt} is a string-theoretical top-down approach to large-$N_c$ QCD. It is based on a proposal for a holographic dual of a non-supersymmetric large-$N_c$ Yang-Mills theory in four effective dimensions by Witten \cite{Witten:1998zw}. In contrast to the 
original AdS/CFT correspondence, the background is provided by the gravitational field of a stack of D4-branes. The dual field theory
now is $4+1$-dimensional since this is the dimension of the world volume of the D4-branes. The extra dimension is compactified on an $S^1$
and thus breaks supersymmetry on the field theory side: by imposing anti-periodic boundary conditions on the adjoint fermions, they obtain a mass of 
the order of the inverse radius of the $S^1$, called Kaluza--Klein mass $M_{\rm KK}$. At one loop level, also the adjoint scalars become massive.
Hence, by choosing the radius of the extra dimension small enough and by restricting to low energies, 
one effectively breaks supersymmetry and effectively reduces the number of dimensions to $3+1$. However, there is a price to 
pay for introducing the extra dimension: 
in order to justify the supergravity approximation for the D4-brane background, the five-dimensional (dimensionful) \mbox{'t Hooft} 
coupling $\lambda_5$ has to be large compared to $M_{\rm KK}^{-1}$. This corresponds to a large four-dimensional
(dimensionless) 't Hooft coupling $\lambda = \lambda_5/(2\pi M_{\rm KK}^{-1})$. In this case, however, the mass gap of the field theory is of the same 
order as $M_{\rm KK}$ and thus the Kaluza-Klein modes do not decouple. Only in the opposite limit $\lambda\ll 1$, where string corrections are 
important and which thus is inaccessible, the Kaluza-Klein modes do decouple and the theory becomes dual to large-$N_c$ QCD in 3+1 dimensions 
(at small energies below the Kaluza-Klein scale). 
It has nevertheless turned out that the classical gravity limit of the D4-brane background is a remarkably useful tool for understanding
certain nonperturbative properties of (large-$N_c$) QCD.

An important property of the Sakai-Sugimoto model is the existence of a Haw\-king--Page transition between a soft-wall and a black hole 
background, which encodes a confinement-deconfinement transition. This feature can be understood either 
from power counting in $N_c$
of the corresponding thermodynamic potentials of the gravity backgrounds or by 
studying the dual to the Wilson line.
Confined and deconfined phases correspond to two different geometric backgrounds which are, in coordinates made 
dimensionless by dividing by the curvature radius $R$, given by
\begin{eqnarray}
\frac{ \D s^2}{R^2}&&=u^{3/2}\left[-h_d(u) \D t^2 +\delta_{ij}\D x^i\D x^j+h_c(u)\D x_4^2\right]
+\frac{\D u^2}{f(u)u^{3/2}}+u^{1/2}\D \Omega_4^2\, , 
\end{eqnarray}
where
\begin{eqnarray}
 f(u)=\left\{\begin{array}{c}
               1-\displaystyle{\frac{u_{\rm KK}^3}{u^3}}\\[1.5ex]
		1-\displaystyle{\frac{u_{T}^3}{u^3}}\end{array}\right. \,  , \quad 
h_d(u)=\left\{\begin{array}{c}
               1\\[1.5ex]
		1-\displaystyle{\frac{u_{T}^3}{u^3}}\end{array}\right. \, , \quad 
h_c(u)=\left\{\begin{array}{cc}
               1-\displaystyle{\frac{u_{\rm KK}^3}{u^3}} & \mathrm{conf.}\\[2ex]
		1 & \mathrm{deconf.}\end{array}\right.  
\end{eqnarray}
and
\begin{eqnarray}
 u_{\rm KK}&&=\left(\frac{4\pi}{3}\right)^2\frac{R^2}{\beta_{x_4}^2}=\frac{4}{9}R^2M_{\mathrm{KK}}^2 \, , 
\qquad  u_T=\left(\frac{4\pi}{3}\right)^2\frac{R^2}{\beta_\tau^2}  \, .
\end{eqnarray}
Here, $\beta_{x_4}$ is the period of $x_4$ -- the coordinate of the additional $S^1$ -- necessary to prevent a conical singularity at $u=u_{\rm KK}$ in the confined phase. The curvature radius is related to the Yang--Mills coupling $g_{\rm YM}$ by 
\begin{eqnarray}
R^3=\pi g_s N_c \ell_s^3 =\frac{g_{\rm YM}^2N_c\alpha'}{2M_{\rm KK}}  \, ,
\end{eqnarray}
where $\ell_s^2 = \alpha'$ is the squared string length. 
In the analytic continuation to Euclidean signature, time is  
also compactified to a circle with circumference $\beta_\tau=T^{-1}$, analogously to finite temperature field theory. 
Increasing the temperature shrinks the Euclidean time circle. At the point where the circumference of the time circle and the extra dimensional circle match, the Hawking--Page transition takes place.  Apart from the metric field the Witten model also contains a nontrivial dilaton and Ramond--Ramond (RR) flux background given by
\begin{equation}
 e^\Phi=u^{3/4}g_s \, , \qquad   F_4=\frac{(2\pi)^3 \ell_s^3 N_c}{\Omega_4}\D\Omega_4 \, , 
\end{equation}
where $\Omega_4$ is the volume of the 4-sphere.

Sakai and Sugimoto introduced fundamental quarks by placing two stacks of $N_f$ D8-branes with opposite orientation into the background in 
the so-called probe limit $N_f\ll N_c$, i.e., back-reactions on the geometry are neglected. In the asymptotic region $u\rightarrow \infty$ the two 
stacks of D-branes are separated on the Kaluza--Klein circle. In the original model they reside at antipodal points. In the bulk, the 
D-branes are space filling in the field theory directions, $x^\mu$, as well as in the $S^4$ and are specified by an embedding function in the 
$u$-$x_4$ subspace. Before going to the gravity description of the D4-branes one can interpret the underlying string picture as follows: strings connecting the D4 with the D8-branes carry one flavor and one color index, hence representing (massless) quarks in the fundamental representation, whereas strings stretching between D8-branes represent mesons. The local symmetry of the $U(N_f)\times U(N_f)$ gauge theory supported on the world volume of the stacks of D8-branes translates into a global symmetry via the holographic dictionary, which is interpreted as the chiral symmetry of the field theory. In the confined background, the two stacks of D8-branes are forced to join at $u_{\rm KK}$ where the additional $S^1$ degenerates and therefore form a single stack with gauge symmetry $U(N_f)$. On the field theory side, this reflects the chiral symmetry breaking mechanism. One can
use a diagonal subgroup of the full symmetry group to introduce chemical potentials and electromagnetic quantities such as an external, 
non-dynamical magnetic field.
Usually the gauge is chosen such that for example the asymptotic value of the zeroth component of the Abelian gauge field is identified with the quark chemical potential. Due to the probe limit, the deconfinement transition is not affected by a finite chemical potential, trivially leading to a phase diagram in the plane $T$-$\mu$ similar to the one discussed for large-$N_c$ QCD in \cite{McLerran:2007qj}.

The low-energy effective theory describing the open string fluctuations is a non--Abelian Dirac--Born--Infeld (DBI) theory on the probe branes;
calculating the fluctuations of the gauge field corresponds to calculating the meson spectrum. Indeed, after fitting the 
value of the 't Hooft 
coupling $\lambda$ and $M_{\rm KK}$ to the rho meson mass and the pion decay constant, the spectrum matches experimental data nicely. 
The mode expansion
used in the calculation of the meson spectrum can also be used to link the Sakai--Sugimoto model to the Skyrme model. 
Apart from the DBI action, the dynamics of D8-branes in a 
background with nontrivial RR-flux is governed by a Chern--Simons (CS) action, since the D8-brane is magnetically charged under that flux. This contribution allows for introducing baryons in the model and is related to chiral solitons in the Skyrme model. Therefore, the full action reads
\begin{eqnarray}
S=&&S_{\mathrm{DBI}}+S_{\mathrm{CS}} \nonumber\\
=&&T_8\int_{D8}\D\tau\D^{8}x\ e^{-\Phi}\Tr\sqrt{|\det(g_{mn}+2\pi\alpha'\mathcal{F}_{mn})|}+\frac{T_8}{6}\int_{D8}C_3\Tr(2\pi\alpha'\mathcal{F})^3
\, ,\quad
\end{eqnarray}
where $T_8$ is the D8-brane tension, and $\D C_3=F_4$. 
Usually one integrates the last term by parts, omitting all boundary terms, to obtain a gauge-variant action 
where the RR-flux couples to the Chern--Simons 5-form.

\begin{figure}[!h]
\includegraphics[width=11cm]{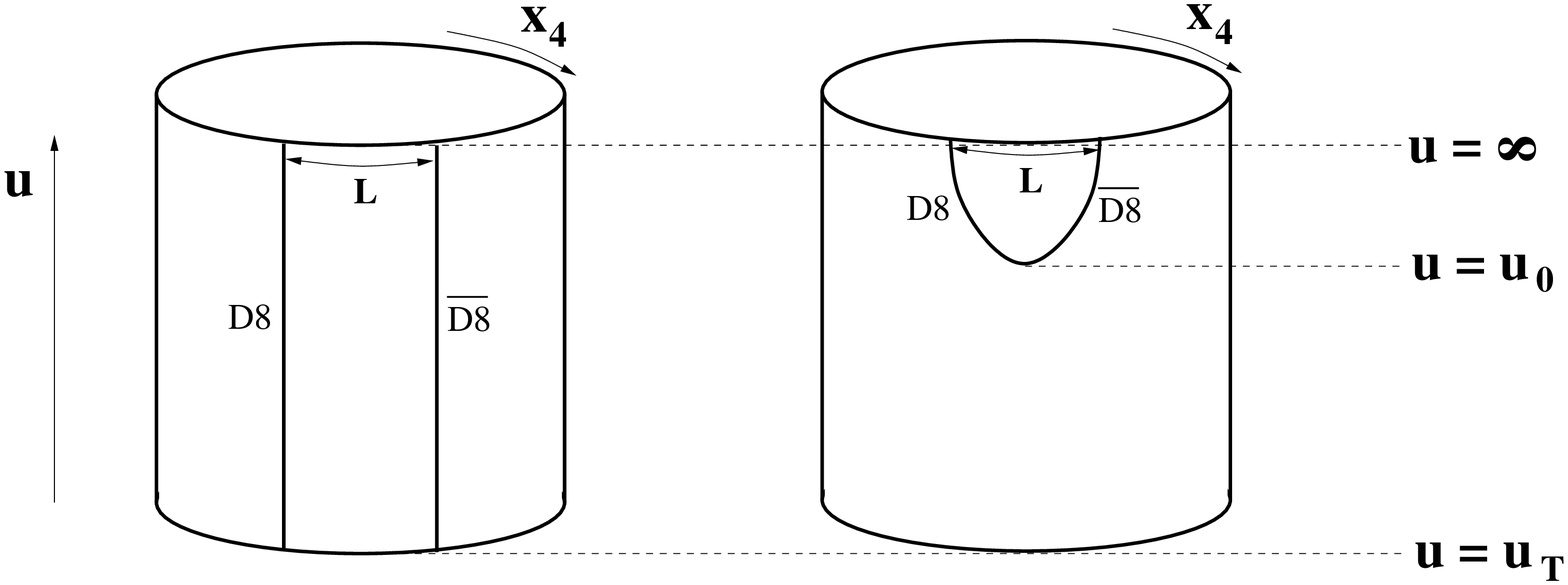}
\caption{The chirally restored (left) and chirally broken (right) phases of the non-antipodal Sakai--Sugimoto model in the deconfined background. 
The calculation reviewed here determines which of the two D8-brane embeddings is favored as a function of temperature, chemical potential, and 
magnetic field for a small asymptotic separation $L$. Only in that limit (in which the dual field theory resembles the NJL model) does the 
chiral phase transition in the probe brane approximation depend on chemical potential and magnetic field.}
\label{zylindersandcigars}
\end{figure}

The Sakai--Sugimoto model also has a connection to the NJL model. In the ``decompactified'' limit where the asymptotic coordinate distance 
between the D8- and anti-D8-branes is much 
smaller than the radius of the extra compactified dimension, $L\ll M_{\rm KK}^{-1}$, the Sakai--Sugimoto model is dual to a non-local NJL 
model \cite{Antonyan:2006vw}.  As a consequence, in the scenario with broken chiral symmetry, the D8-branes now in 
general join at $u_0>u_{\rm KK}$. The difference $u_0-u_{\rm KK}$ is commonly interpreted as the constituent quark mass within a meson, which is 
realized as a string with both end points attached to the tip of the joined D8-branes hanging down to the bottom of the geometry. 
With a sufficiently small asymptotic separation of the flavor branes, it is also possible to find an energetically preferred phase 
with broken chiral symmetry in the deconfined background \cite{Aharony:2006da}, see Fig.\ \ref{zylindersandcigars}. The resulting phase 
diagram at finite chemical potential was first discussed in \cite{Horigome:2006xu}. By reducing $L$ compared to $M_{\rm KK}^{-1}$, the temperature 
range 
where the system is confined becomes small compared to the temperature range governed by the deconfined and chirally 
broken phase. Eventually, the resulting phase diagram resembles the NJL result (where no confined phase is present) 
shown in the right panel of Fig.\ \ref{finiteT}. Consequently, the Sakai--Sugimoto 
model allows for interpolating between a non-local NJL model ($L\ll M_{\rm KK}^{-1}$) and -- modulo the above mentioned caveats -- large-$N_c$ 
QCD ($L = \pi M_{\rm KK}^{-1}$). In the former limit, the flavor D8-branes do not probe deeply the background geometry produced by the color
D4-branes (which corresponds to neglecting gluon dynamics), while in the latter the gluons dominate.   

The effect of a homogeneous background magnetic field has first been considered in \cite{Bergman:2008sg}. Shortly thereafter, the effect on 
the critical temperature for chiral symmetry restoration at vanishing chemical potential has been analyzed \cite{Johnson:2008vna}. Like
in the NJL result from Fig.\ \ref{gapTmueq0}, the critical temperature increases with the magnetic field, which shows that the Sakai-Sugimoto
model exhibits magnetic catalysis. Finite chemical potentials have been introduced together with a magnetic field in \cite{Thompson:2008qw} 
in the original Sakai--Sugimoto model. The deconfined, chirally symmetric phase was discussed in \cite{Lifschytz:2009sz}, where a magnetic phase 
transition within the symmetric phase was found that is reminiscent of a transition to the lowest Landau level. 
The full phase diagram in the parameter space $T$-$\mu$-$B$ in the 
deconfined phase was presented in our work \cite{Preis:2010cq}. In particular, the effect of inverse magnetic catalysis effect was 
found and discussed in this reference. 

The Sakai-Sugimoto model can be developed further to include homogeneous baryonic matter, made from 
point-like approximations to the solitonic baryons mentioned above \cite{Bergman:2007wp}. Applications in the context of a background magnetic field 
have been studied in the confined \cite{Bergman:2008qv} and deconfined \cite{Preis:2011sp} backgrounds, the latter study investigating the effect 
of baryonic matter on inverse magnetic catalysis. Here we shall mostly focus on the case without baryons, and only at the end of 
Sec.\ \ref{sec:chiralSS} briefly review their effect on the phase diagram.

\subsection{Equations of motion and axial current}

In terms of an embedding function $x_4(u)$ for the D8-branes, the induced metric reads
\begin{equation}
\frac{ \D s_{D8}^2}{R^2}=
u^{3/2}h_d \D t^2 +u^{3/2}\delta_{ij}\D x^i\D x^j+u^{3/2}\left(x_4'^2 h_c+\frac{1}{fu^3}\right)\D u^2+u^{1/2}\D\Omega_4^2 \, ,
\end{equation}
where the prime denotes derivative with respect to $u$. We work with one flavor, $N_f=1$, and for the (dimensionless) $U(1)$ gauge field we choose the ansatz
\begin{equation} \label{afield}
a=\frac{2\pi\alpha'}{R}\mathcal{A}_\mu\D x^\mu=a_0(u)\D t+bx^1\D x^2+a_3(u)\D x^3,
\end{equation}
where $b=2\pi\alpha'B$ denotes the magnitude of the dimensionless magnetic field strength. Note that the necessity of introducing the third 
component of the gauge field, which is P-odd, is due to the coupling to $a_0$ and $b$ via the (P-odd) CS-action. We denote the asymptotic values of 
the gauge field by 
\begin{eqnarray}
a_0(\infty)=\mu \equiv \mu_q\frac{2\pi\alpha'}{R} \, , \qquad a_3(\infty)=j \, , \qquad x_4(\infty)=\frac{\ell}{2} \, , 
\end{eqnarray}
where $\mu_q$ is the dimensionful quark chemical potential\footnote{Here we keep the notation of Refs.\ \cite{Preis:2010cq,Preis:2011sp}. 
Note that in the NJL section $\mu$ is the dimensionful quark chemical potential.}, and $\ell\equiv L/R$ is the dimensionless asymptotic 
separation of the flavor 
branes. The boundary value of $a_3$ can be shown to correspond to a finite expectation value for the pion 
gradient in the direction of the magnetic field, hence it will only be nonvanishing when chiral symmetry is broken. In that case, one 
has to extremize the on-shell action with respect to $j$ \cite{Thompson:2008qw,Bergman:2008qv,Rebhan:2008ur}. 
From the field theory perspective this means that, if $j\neq 0$, the chiral condensate 
is rotating between a scalar and a pseudoscalar condensate when moving along the $z$-direction, i.e., it forms a so-called chiral spiral 
\cite{Schon:2000he}. Each full turn of the spiral raises the baryon number by one. Therefore, since $j$ measures the rate of turns per 
unit length, it is related to the baryon density. Equivalently, one can regard $j$ as a supercurrent, in analogy to superfluidity, where the
phase of the condensate gives the superfluid velocity.

Before continuing we put some restrictions on the gauge field and the embedding: in the joined configuration we assume that the fields are continuous at the junction point $u_0$ since for now we omit any point-like sources, hence $a_3(u_0)=0$ and $x_4'(u_0)=\infty$. In the restored phase, 
due to the presence of the horizon, we have to satisfy the regularity constraint $a_0(u_T)=0$ \cite{Horigome:2006xu}.

Within our ansatz, the action for the D8-brane describing left-handed fermions becomes
\begin{eqnarray}
S'= \frac{\mathcal{N}V}{2T}\int_{u_0/u_T}^\infty \D u&&\left[\sqrt{u^5+b^2u^2}\sqrt{u^3fx_4'^2+\frac{h_d}{f}-a_0'^2+a_3'^2h_d}+\right.\nonumber\\
&&\left.+ \frac{3b}{2}(a_3a_0'-a_0a_3')\right]\, , \label{action1}
\end{eqnarray}
where the lower bound of the integration has to be chosen according to the phase under consideration and where  
\begin{eqnarray}
 \mathcal{N}\equiv 2\frac{T_8R^5\Omega_4}{g_s}=&&\frac{N_cR^2}{6\pi^2(2\pi\alpha')^3} \, .
\end{eqnarray}
Here we have modified the original action $S$ and denoted the new action by $S'$. The reason is that proceeding with $S$ results in an inconsistency: 
the conserved currents sourced by the boundary values of the gauge field turn out to be different from those currents calculated using the thermodynamic relations. In \cite{Bergman:2008qv} this issue was related to the gauge variance of the CS action. The solution to this problem is to 
supplement the CS action with boundary terms residing at the holographic as well as at the spatial boundaries. After integration by parts this 
modification amounts to simply multiplying the CS contribution with a factor $3/2$.

The integrated equations of motion are
\begin{eqnarray}
\frac{\sqrt{u^{5}+b^2u^2}a_0'}{\sqrt{u^3fx_4'^2+\frac{h_d}{f}-a_0'^2+a_3'^2h_d}}&&=3ba_3+c \, ,
\label{a0eom1}\\
\frac{\sqrt{u^{5}+b^2u^2}h_da_3'}{\sqrt{u^3fx_4'^2+\frac{h_d}{f}-a_0'^2+a_3'^2h_d}}&&=3ba_0+d \, , \label{a3eom1}\\
\frac{\sqrt{u^5+b^2u^2}fu^3x_4'}{\sqrt{u^3fx_4'^2+\frac{h_d}{f}-a_0'^2+a_3'^2h_d}}&&=k \, .\label{x4eom1}
\end{eqnarray}
The left-hand side of Eq.\ (\ref{a0eom1}) is the magnitude of the (bulk) electric field corresponding to the gradient of $a_0$ in a curved 
background on one D8-brane pointing towards larger values of $u$. When we move past the point $u_0$ in the joined D-brane configuration the direction of the electric field is flipped since we assume that $a_0$ is P-even. Therefore, since $a_3(u_0)=0$, the integration constant $c$ corresponds to a point-like 
source at $u_0$. For now, we do not include any point-like baryons and thus set $c=0$ in the broken phase. In the restored phase, 
on the other hand, $c\neq 0$, hence the horizon provides a charge that will be translated into the quark density at the boundary. (In the 
restored phase, $x_4'(u)\equiv 0$ and thus $k=0$, i.e., only two nontrivial equations remain.) 
Furthermore, if the magnetic field is nonzero there is an additional contribution to the quark density from the 
gradient of $a_3$, which in general is distributed over the whole D8-brane world volume. Equation (\ref{a3eom1}) evaluated at $u_T$ enforces us to set $d=0$ in the restored phase in order to maintain consistency  since $h_d(u_T)=a_0(u_T)=0$.

The nonvanishing components of the current densities sourced by the asymptotic gauge field components are given by
\begin{eqnarray}
 \mathcal{J}^0_V= \mathcal{J}^0_R+ \mathcal{J}^0_L =\frac{2\pi\alpha'\mathcal{N}}{R}\left(\frac{3b}{2}j+c\right)\, ,\label{J0V}\\
 \mathcal{J}^3_A= \mathcal{J}^3_R- \mathcal{J}^3_L=\frac{2\pi\alpha'\mathcal{N}}{R}\left(\frac{3b}{2}\mu+d\right)\, ,
\end{eqnarray}
where we have used the equations of motion. The first line relates the baryon density with
the \textit{magnetic} chiral spiral and the point-like charges in the bulk. The second line is the axial current which we have already encountered in section \ref{sec:MC}. Because we have to extremize the thermodynamic potential with respect to $j$, i.e., with respect to $a_3(\infty)$, we can immediately conclude that in the broken phase the axial current has to vanish, hence $d=-3/2\ b\mu$. In the chirally symmetric phase the axial current at \textit{any temperature} -- reinstating dimensionful quantities -- reads
\begin{equation}
 \mathcal{J}^3_A=\frac{N_c}{4\pi^2}B\mu_q \, .
\end{equation}
This result differs from the corresponding expression (\ref{NJLaxialcurrent}) obtained in the NJL model by a factor $2$, which is related 
to the modification of the CS term in the action in order to obtain a consistent thermodynamic description of the currents. 
For a thorough discussion of the effect of this modification on the chiral anomaly see Ref.\ \cite{Rebhan:2009vc}.

\subsection{Semianalytic solution to the equations of motion}

In general, from this point on one has to rely on numerical methods. However, using $f(u)\simeq 1$ we can go a little further. 
This approximation is valid either in the deconfining background if $T=0$ or in the decompactified limit of the confined background. Moreover, as will be justified a posteriori, for $L\ll M_{\rm KK}^{-1}$ and in the chirally broken phase we have $u_0\gg u_{\rm KK}$ (confined) or $u_0\gg u_{T}$ 
for sufficiently small $T$ (deconfined). We will later work in the deconfined geometry and apply this approximation for the chirally broken phase 
at any $T$ (i.e., our approximation becomes less accurate for large $T$). If chiral symmetry is restored this is of course not allowed, since the D8-branes extend from the holographic boundary down to the horizon at $u_T$. Hence, when computing the phase diagram we will compare the grand canonical potential of the broken phase using the $f(u)\simeq 1$ approximation with the full numerical result obtained for the restored phase. Note that in the special case $b=0$ or $\mu=0$ the temperature can be easily introduced in the symmetric phase since the ``blackening" function $f(u)$ does not appear explicitly in the equations of motion. There temperature enters only in the lower bound $u_T$ of the integrals over the holographic coordinate.

With $f(u)\simeq 1$, 
we can simplify Eqs.\ (\ref{a0eom1}) and (\ref{a3eom1}) considerably. We define the new coordinate field $y(u)$ via the differential 
equation
\begin{equation}
y'=\frac{3bu^{3/2}}{\sqrt{u^8+u^5b^2-k^2+(3b)^2u^3[(\partial_y a_0)^2-(\partial_y a_3)^2]}} \, , \label{eomfory}
\end{equation}
for which we have the freedom to choose $y(u_0)=0$ or $y(0)=0$ in the broken and symmetric phase respectively. Its value at the holographic boundary will be denoted by $y_\infty$ in the following. In the joined D8-brane configuration the boundary condition $x_4'(u_0)\rightarrow \infty$ implies that $y'(u_0)\rightarrow \infty$.
After algebraically rearranging Eqs.\ (\ref{a0eom1}) - (\ref{x4eom1}) such that all derivatives with respect to $u$ are placed on the left-hand 
side, the equations of motion for the gauge fields as a function of the new coordinate $y$ are 
\begin{eqnarray}
\partial_y a_0&&=a_3+\frac{c}{3b}\,, \qquad \partial_y a_3=a_0+\frac{d}{3b} \, ,
\end{eqnarray}
for which we can easily find the solutions
\begin{eqnarray}
a_0&&=c_1\cosh y +c_2\sinh y -\frac{d}{3b} \, ,\qquad a_3=c_1\sinh y +c_2\cosh y -\frac{c}{3b}.
\end{eqnarray}
This allows us to write the grand canonical potential, i.e., the on-shell action,  as
\begin{equation}
\Omega=\mathcal{N}\left[\int_{u_0/u_T}^\infty\frac{3b}{y'}\D u +\frac{kL}{2}-\frac{3b}{2}y_\infty\left(c_2^2-c_1^2\right)-\frac{c}{2}(\mu-c_1)+\frac{d}{2}\left(j-c_2\right)\right] \, .
\end{equation}
This expression is divergent. 
In order to obtain finite expressions we renormalize the grand canonical 
potential by the chirally symmetric vacuum contribution
\begin{eqnarray} \label{vacuumOm}
\Omega(\mu=T=0)&=&\mathcal{N}\int_0^\Lambda \D u \sqrt{u^5+b^2u^2} \, .
\end{eqnarray}
The integration constants found by imposing the boundary conditions discussed below Eq.\ (\ref{afield}) and the supercurrent $j=a_3(\infty)$
are summarized in table \ref{intconst}. 
\begin{table}[t]
\begin{center}
\begin{tabular}{| c ||c| c|c|c|c|c|}
\hline
\rule[-1.5ex]{0em}{4ex}
 & $d$ & $c_1$  & $c_2$ & $c$ & $k$ & $j$ \\
\hline\hline
\rule[-1.5ex]{0em}{6ex}
 broken & $\;\;-\frac{3}{2}\mu\;\;$ & $\;\;\frac{\mu}{2\cosh y_\infty}\;\;$ & $0$ & $0$ & $\;\;\sqrt{u_0^8+b^2u_0^5-\left(\frac{3b\mu}{2\cosh y_\infty}\right)^2u_0^3}\;\;$ & $\;\;\frac{\mu}{2}\tanh y_\infty\;\;$
\\[2ex]
\hline
\rule[-1.5ex]{0em}{5ex}
  restored & $0$ & $0$ & $\;\;\frac{\mu}{\sinh y_\infty}\;\;$ & $\;\;3b\mu\coth y_\infty\;\;$ & $0$ & 0 \\[2ex]
\hline
\end{tabular}
\caption{The integration constants $d$, $c_1$, $c_2$, $c$, $k$ and the supercurrent $j$ for the chirally broken and restored phases.
\label{intconst}}
\end{center}
\end{table}

\subsection{Broken chiral symmetry}

Inserting the supercurrent $j$ and the constant $c$ from table \ref{intconst} into Eq.\ (\ref{J0V}) yields  the quark number density 
\begin{equation}
n_q\equiv \mathcal{J}^0_V=\frac{N_c}{8\pi^2}B\mu_q\tanh y_\infty \, .
\end{equation}
The only equations that remain and in general have to be solved numerically for the variables $u_0$ and $y_\infty$ are
\begin{eqnarray}
\frac{\ell}{2}&&=\sqrt{u_0^8+b^2u_0^5-\left(\frac{3b\mu}{2\cosh y_\infty}\right)^2}\int_{u_0}^\infty\frac{\D u}{u^{3/2}g(u)} \, , \quad
y_\infty =3b\int_{u_0}^\infty\frac{u^{3/2}\D u}{g(u)} \, ,
\end{eqnarray}
where we have abbreviated
\begin{eqnarray}
g(u)\equiv \sqrt{u^8+b^2u^5-\left(\frac{3b\mu}{2\cosh y_\infty}\right)^2 u^3-u_0^8-b^2u_0^5-\left(\frac{3b\mu}{2\cosh y_\infty}\right)^2u_0^3} \, .
\end{eqnarray}
Note that the explicit dependence on the asymptotic separation $\ell$ can be eliminated by the rescaling $u\rightarrow \ell^2u$, 
$\mu\rightarrow \ell^2\mu$, $b\rightarrow \ell^3b$ and $\Omega\rightarrow \ell^7\Omega$. Therefore, in all plots shown below, 
the axes are measured in appropriate units of the D8-brane separation.

Before coming to the full numerical results, let us first discuss the two limits of small and large magnetic fields $b$. 
For a detailed derivation of the approximations consult appendix D in Ref.\ \cite{Preis:2010cq}. 

For small magnetic fields, $y_\infty$ and thus the supercurrent $j$ rise linearly with $b$, and therefore the lowest order contribution 
to the quark number density induced by the chiral spiral is quadratic in $b$.
The location of the tip of the connected flavor branes is $u_0 \simeq u_0^{(0)}+ \eta_1(\mu)b^2$ with the value of $u_0$ at $b=0$,
\begin{equation}
u_0^{(0)}= \left[\frac{4\sqrt{\pi}\Gamma\left(\frac{9}{16}\right)}{\ell\Gamma\left(\frac{1}{16}\right)}\right]^2\simeq 0.5249\ \ell^{-2} \, .
\end{equation}
Interestingly, the $\mu$-dependent coefficient $\eta_1$ possesses a zero at 
$\mu\simeq 0.2905/\ell^2$, above which it becomes negative. This shows that the constituent quark mass (which is given by $u_0$) 
can {\it decrease} with the magnetic field for sufficiently large chemical potentials. This behavior can be traced back to the incorporation of 
the chiral spiral.

The grand canonical potential (renormalized by the vacuum contribution (\ref{vacuumOm})) is approximated for small $b$ by
\begin{eqnarray} \label{Ombroken1}
\Omega_{\mathrm{ren}}&&\simeq -\mathcal{N}\left[\frac{2}{7}(u_0^{(0)})^{7/2}\frac{\sqrt{\pi}\Gamma\left(\frac{9}{16}\right)}{\Gamma\left(\frac{1}{16}\right)}+\eta_2(\mu)b^2\right]\, , 
\end{eqnarray}
where
\begin{eqnarray}
\eta_2(\mu)\equiv\frac{\sqrt{\pi}\Gamma\left(\frac{9}{16}\right)}{\Gamma\left(\frac{1}{16}\right)}\sqrt{u_0^{(0)}}\left[\cot\frac{\pi}{16}+\left(\frac{3\mu}{2u_0^{(0)}}\right)^2\frac{\Gamma\left(\frac{3}{16}\right)\Gamma\left(\frac{17}{16}\right)}{\Gamma\left(\frac{9}{16}\right)\Gamma\left(\frac{11}{16}\right)}\right] \, .
\end{eqnarray}
(As explained in \cite{Preis:2010cq}, there exists a second solution in the region of small $b$, where $u_0$ is small and $y_\infty$ is large, 
which is separated from the solution discussed here by a first order phase transition. However, this first-order phase transition
occurs in a region of large $\mu$ where the chirally restored phase is preferred. Therefore we will not discuss this second solution here.)

At asymptotically large magnetic field, $y_\infty$ diverges faster than linearly, thus $j\simeq\mu/2$, while $u_0$ saturates at the value
\begin{equation}
u_0^{(\infty)} = \left[\frac{4\sqrt{\pi}\Gamma\left(\frac{3}{5}\right)}{\ell\Gamma\left(\frac{1}{10}\right)}\right]^2\simeq 1.2317\ \ell^{-2}.
\end{equation}
We see that $u_0^{(\infty)}>u_0^{(0)}$, i.e., for any $\mu$ the constituent quark mass at asymptotically large $b$ is larger than that at $b=0$.
This can be interpreted as magnetic catalysis and is similar to the NJL model. However, as we have shown in the left panel of 
Fig.\ \ref{gapTmueq0}, in the NJL model the constituent quark mass does not saturate for asymptotically large  magnetic fields.

Plugging these results into $\Omega$ and $n_q$ yields
\begin{eqnarray} \label{Omegabroken2}
\Omega_{\mathrm{ren}}&&\simeq -\mathcal{N}b\left[\frac{\sqrt{\pi}\Gamma\left(\frac{3}{5}\right)}{2\Gamma\left(\frac{1}{10}\right)}
(u_0^{(\infty)})^2+\frac{3\mu^2}{8}\right] \, ,\qquad n_q \simeq \frac{N_c}{8\pi^2}B\mu_q.
\end{eqnarray}
Remarkably, all model parameters have dropped out of the quark number density, which thus is solely expressed in terms of the dimensionful
quantities $B$ and $\mu_q$. 

\subsection{Symmetric phase}
\label{sec:sym}

The following analytical expressions are all valid in the zero-temperature limit. Only in the plots at the end
of this subsection we include numerical finite-temperature results. 
Now only one equation remains to be solved numerically for $y_\infty$,
\begin{equation}
y_\infty=\int_{0}^\infty\frac{3bu^{3/2}}{\sqrt{u^8+b^2u^5+\left(\frac{3b\mu}{\sinh y_\infty}\right)^2 u^3}}\D u \, .
\end{equation}
For $b>0$, this equation has in general three solutions: $y_\infty=\infty$, which is always a solution, and two finite solutions, the larger 
of which turns out to be unstable. At sufficiently large values of $b$ for a given $\mu$ only the divergent solution survives. For 
the quark density we find 
\begin{eqnarray}
n_q&&=\frac{N_c}{2\pi^2}B\mu_q\coth y_\infty \, .
\end{eqnarray}
Let us first take the limit where $b$ is small. In this case, $y_\infty$ is linear in $b$, and we obtain for the (dimensionful) quark number density
\begin{equation}
n_q=\frac{\sqrt{N_cM_{\rm KK}}}{3g_{YM}\pi^{3/2}}\mu_q^{5/2}\left[\frac{\sqrt{\pi}}{\Gamma\left(\frac{3}{10}\right)\Gamma\left(\frac{6}{5}\right)}\right]^{5/2} +\mathcal{O}(B^2) \,  .
\end{equation}
The unusual exponent $5/2$ of $\mu_q$ can only occur due to the presence of the dimensionful model parameter $M_{\rm KK}$ (due to  
the extra dimension in the model), which provides the 
missing mass dimensions. 

The grand canonical potential becomes for small $b$
\begin{eqnarray}
\Omega_{\mathrm{ren}}&&\simeq -\mathcal{N}\left\{\frac{2}{7}\mu^{7/2}\left[\frac{\sqrt{\pi}}{\Gamma\left(\frac{3}{10}\right)\Gamma\left(\frac{6}{5}\right)}\right]^{5/2} + \eta_3 b^2\sqrt{\mu} \right\}\label{OmegahLL} \, , 
\end{eqnarray}
with
\begin{eqnarray}
\eta_3&&\equiv
\frac{3}{2}\left[\frac{\Gamma\left(\frac{3}{10}\right)\Gamma\left(\frac{6}{5}\right)}{\sqrt{\pi}}\right]^{5/2}+\frac{\Gamma\left(\frac{9}{10}\right)\Gamma\left(\frac{3}{5}\right)}{\pi^{1/4}\sqrt{\Gamma\left(\frac{3}{10}\right)\Gamma\left(\frac{6}{5}\right)}} \, .
\end{eqnarray}

\begin{figure}
\begin{minipage}{\textwidth}
\includegraphics[height=3.7cm]{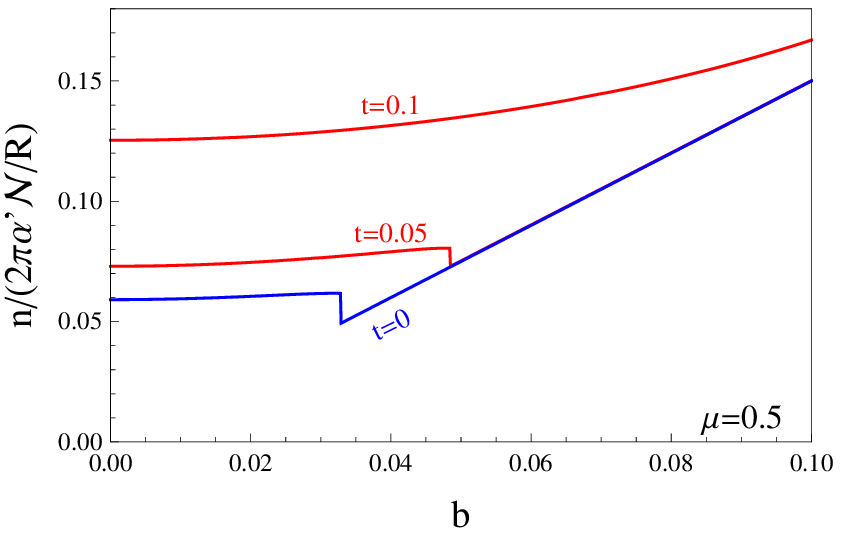}
\hspace{\fill}
\includegraphics[height=3.7cm]{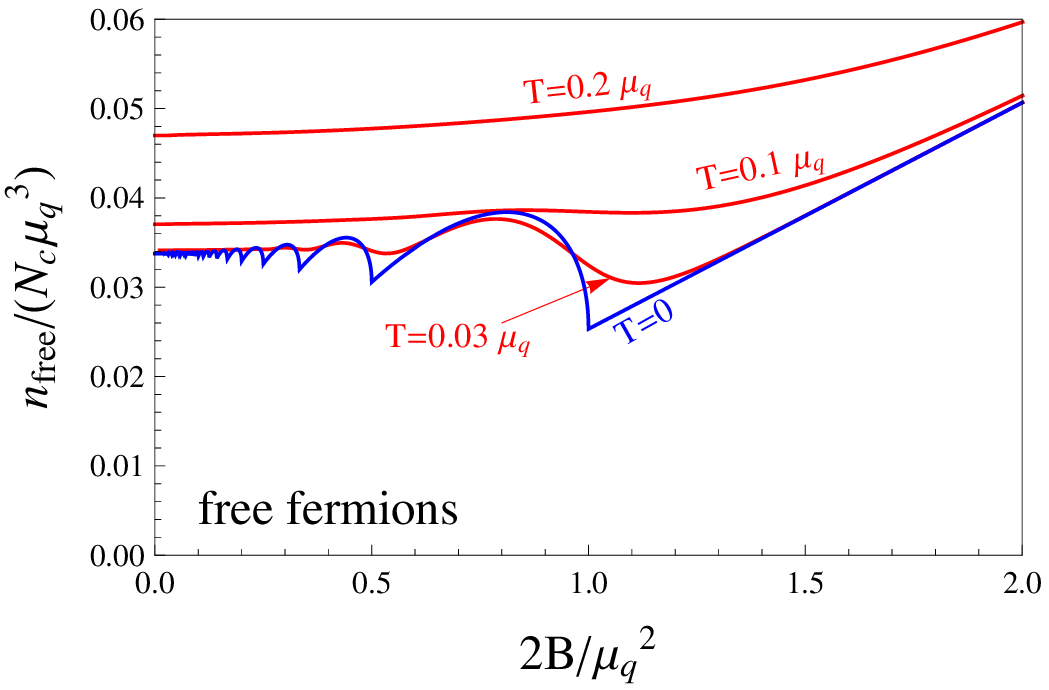}
\end{minipage}
\caption{Quark number density as a function of the background magnetic field for a given chemical potential at various (dimensionless) temperatures 
$t\equiv TR$ in the 
Sakai-Sugimoto model (left) and the NJL model (right).}
\label{densitiesSS}
\end{figure}

Taking the limit $b\rightarrow \infty$ allows only the solution $y_\infty=\infty$, as mentioned before. However, note that this is also a valid solution at finite $b$, hence the following results carry over to any value of $b$ as long as this particular phase is considered. 
Interestingly, the density in this case is
\begin{equation}
n_q=\frac{N_c}{2\pi^2}B\mu_q,
\end{equation}
which takes precisely the form of the density of gapless free fermions in the lowest Landau level. Therefore, we may speak
of a LLL-like phase in the Sakai-Sugimoto model, although there are, because of the strong-coupling nature, no quasiparticles and thus
no Landau levels in the actual sense. 
The grand canonical potential is
\begin{equation}
\Omega_{\mathrm{ren}}=-\mathcal{N}\frac{3b\mu^2}{2}\label{LLLcosts} \, .
\end{equation} 
Using (\ref{OmegahLL}) together with (\ref{LLLcosts}) we can derive the critical magnetic field of the first-order transition within the
chirally restored phase to the LLL--like phase as a function of the chemical potential,
\begin{equation} \label{bcapp}
b_c\simeq 0.095 \mu^{3/2}.
\end{equation}
In the left panel of  Fig.\ \ref{densitiesSS} we plot the quark number density for different temperatures. As a comparison, 
we also plot the corresponding density for (massless) free fermions in a magnetic field, obtained by taking the derivative 
with respect to the chemical potential of the thermodynamic potential (\ref{OmTB}).  

In the case of free fermions,  the higher Landau levels cause oscillations in the density at small magnetic field. 
These oscillations are absent in the ``higher Landau level phase'' in the Sakai-Sugimoto model, given by the solution 
$y_\infty<\infty$. This might be a consequence of the strong coupling, in which case we do not
expect a sharp Fermi surface, even at $T=0$. Furthermore, in the NJL model, the transitions between the phases with differently filled 
Landau levels, in particular also the transition to the LLL phase, is second order, while in the Sakai--Sugimoto model it is first order. 
At finite temperature, the transitions become immediately smooth in the NJL model, while for given $\mu$ it remains first order in the 
Sakai--Sugimoto model until a critical temperature is reached, which increases with increasing $\mu$. 
Above this temperature only one minimizing solution for $y_\infty$ exists for all $b$ and given $\mu$. As a result, the transition line
in the $b$-$\mu$ plane has a critical endpoint for a given temperature, resulting in a critical line in the three-dimensional phase diagram,
see Fig.\ \ref{SSpd}. 
Another important difference is the location of the LLL-transition in the $\mu$-$b$ diagram: the critical magnetic field at zero temperature
is proportional to $\mu^{3/2}$, compared to $\mu^2/2$ for free fermions. Again this is due to the occurrence of $\sqrt{M_{\rm KK}}$.

\subsection{Chiral phase transition}
\label{sec:chiralSS}

First we discuss the critical temperature for chiral symmetry restoration at vanishing chemical potential. In this case, 
in the restored phase the only temperature dependence enters via the lower bound of the integrals over the holographic 
coordinate, $u_T=(4\pi t/3)^2$, with $t=RT$. 
Therefore, one easily determines the renormalized grand canonical potential of the restored phase for the cases $b=0$, 
$\Omega_{\mathrm{ren}}=-2/7{\cal N} u_T^{7/2}$, and   $b\rightarrow \infty$,  $\Omega_{\mathrm{ren}}=-{\cal N} b u_T^{2}/2$.
Then, together with the corresponding expressions for the broken phase from Eqs.\ (\ref{Ombroken1}) and (\ref{Omegabroken2}) we compute the
critical temperatures
\begin{eqnarray}
t_c(\mu=b=0)&&=0.1355/\ell \, , \\
t_c(\mu=,b\rightarrow\infty)&&=0.1923/\ell \, .\label{tcbinfty}
\end{eqnarray}
(Remember that we have used the $f(u)\simeq 1$ approximation for the broken phase which, strictly speaking, is only valid for very small 
temperatures.) We see that the Sakai--Sugimoto model reproduces the usual magnetic catalysis effect at zero chemical potential
because the critical temperature at asymptotically large $b$ is larger than that at vanishing $b$. This is supported by the numerical solution 
which shows that the critical temperature increases monotonically with the magnetic field.
In contrast to the NJL model, the critical temperature saturates at the value given in equation (\ref{tcbinfty}), because the value for $u_0$, 
i.e., the holographic constituent quark mass, saturates.

\begin{figure}[h!]
\begin{minipage}{\textwidth}
\includegraphics[height=5cm]{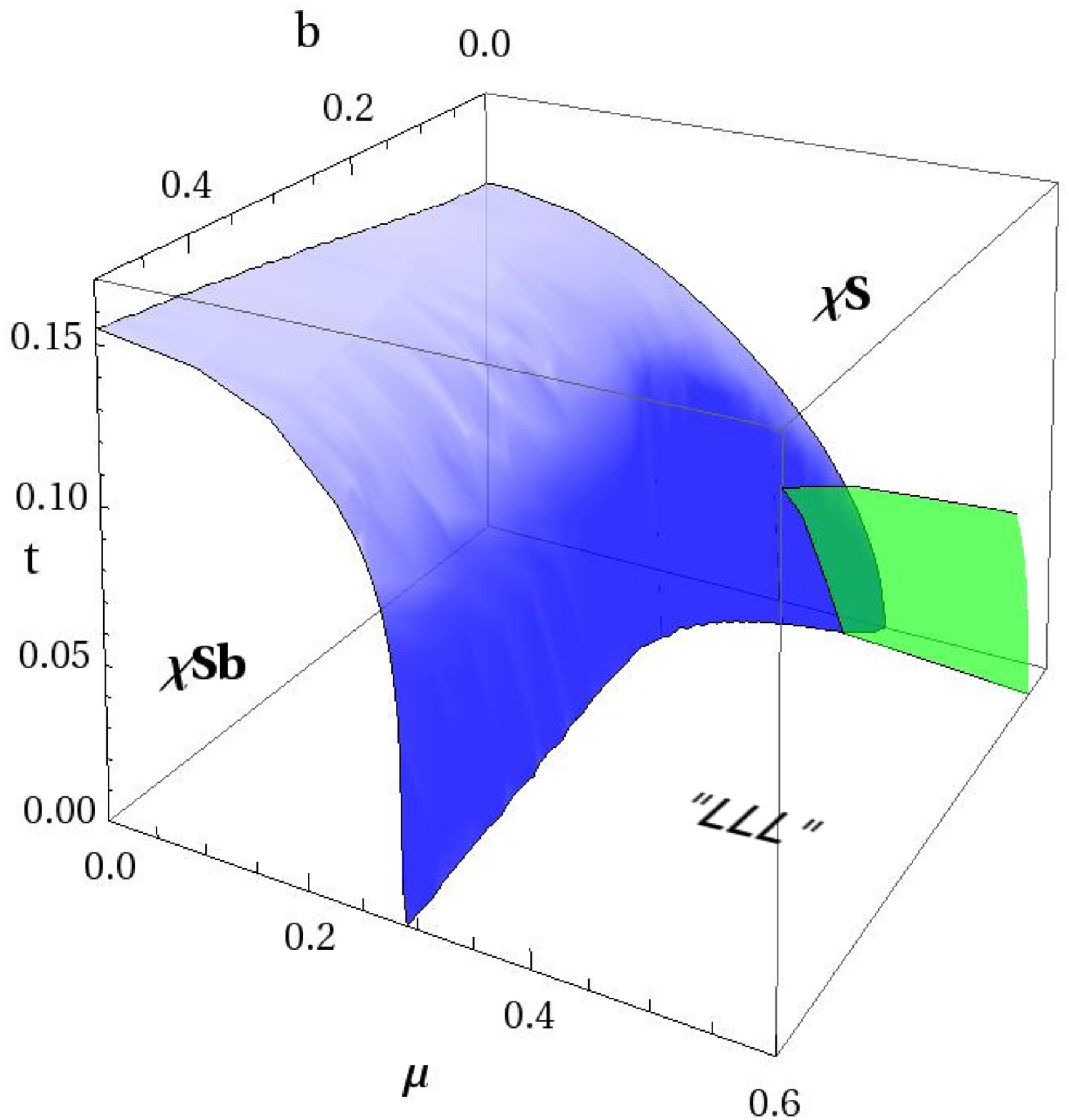}
\hspace{\fill}
\includegraphics[height=3.5cm]{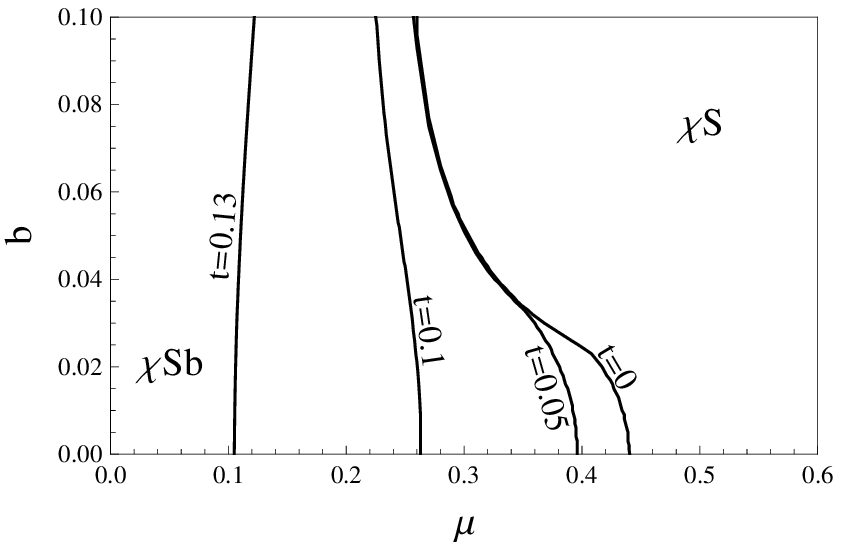}
\includegraphics[height=3.5cm]{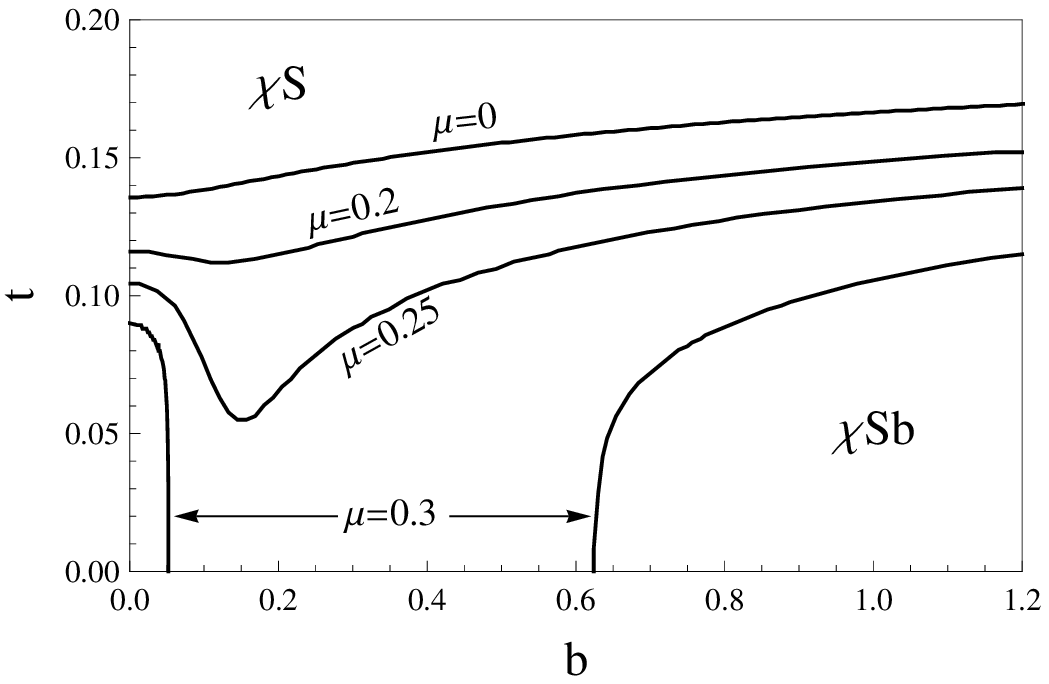}
\hspace{\fill}
\includegraphics[height=3.5cm]{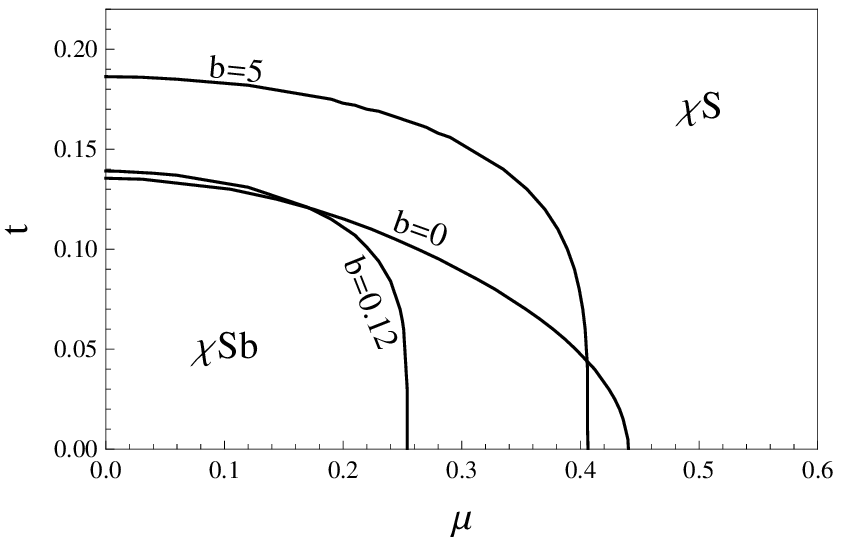}
\end{minipage}
\caption{Upper left panel: The surface of the chiral phase transition (blue) in the deconfined phase of the Sakai--Sugimoto model in the 
$T$-$\mu$-$B$ space. The small (green) surface shows the transition from the ``higher LL" phase to the ``LLL" phase, explained
in Sec.\ \ref{sec:sym}. Upper right, lower left and lower right panels:  two-dimensional cuts at various fixed temperatures, chemical potentials 
and magnetic fields, respectively, through the three-dimensional phase diagram. In the lower left plot, for instance, 
we see that the monotonically increasing critical temperature at $\mu=0$ becomes a non-monotonic curve at finite $\mu$ and may even turn into
two disconnected pieces, separating two chirally broken phases at small and large magnetic fields.}
\label{SSpd}
\end{figure}

At zero temperature, we use Eqs.\ (\ref{Ombroken1}) and (\ref{Omegabroken2}) for the broken phase and Eqs.\ (\ref{OmegahLL}) and 
(\ref{LLLcosts}) 
for the restored phase to 
compute the critical chemical potentials
\begin{eqnarray}
\mu_c(t=b=0)&&=0.4405/\ell^2 \, ,\\
\mu_c(t=0, b\rightarrow\infty)&&=0.4325/\ell^2 \, .
\end{eqnarray}
This result already shows that inverse magnetic catalysis in the sense explained in Sec.\ \ref{sec:IMCNJL} must be present in the 
Sakai--Sugimoto model. The full numerical solution of the surface of the chiral phase transition in the three dimensional $T$-$\mu$-$B$
space, including cuts through the surface at fixed $t$, $\mu$, and $b$, is shown in Fig.\ \ref{SSpd}.

\begin{figure}[h]
\begin{minipage}{\textwidth}
\begin{center}
\includegraphics[width=0.7\textwidth]{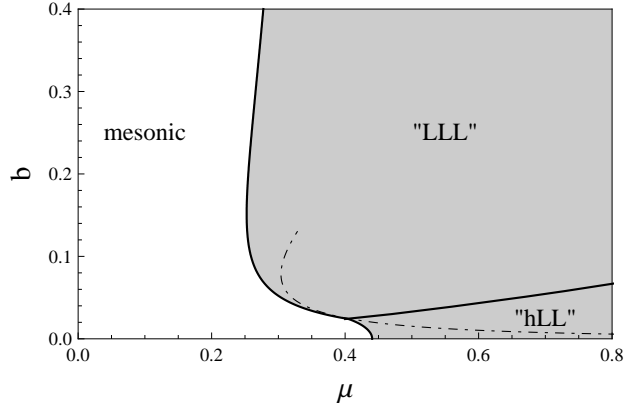}
\end{center}
\end{minipage}
\caption{The chiral phase transition at zero temperature from the Sakai-Sugimoto model (ignoring baryonic matter). The chirally 
broken phase (white) is separated by a first-order phase transition (solid line) from the chirally restored phase (gray). 
The dashed-dotted line is the approximation from Eq.\ (\ref{imcSS}). Translating the dimensionless 
quantities $b$ and $\mu$ into physical units \cite{Preis:2010cq}, one concludes that  
the magnetic field decreases the critical chemical potential from $\mu_q\simeq 400\,{\rm MeV}$ at 
$|qB|=0$ down to $\mu_q\simeq 230\,{\rm MeV}$ at $|qB|\simeq 1.0\times 10^{19}\,{\rm G}$ where the critical line turns around and the 
critical chemical potential starts to increase with $|qB|$.} 
\label{fig:IMCSS}
\end{figure}  
\begin{figure}[h!]
\begin{minipage}{\textwidth}
\begin{center}
\includegraphics[width=0.7\textwidth]{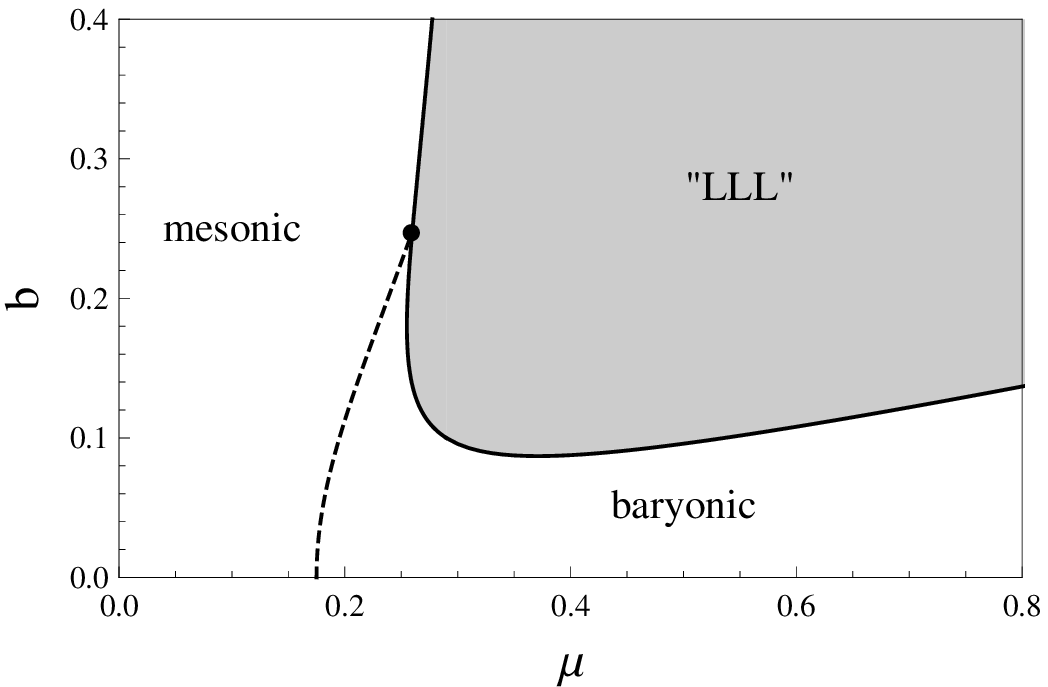}
\end{center}
\end{minipage}
\caption{As Fig.\ \ref{fig:IMCSS}, but including baryonic matter (from Ref.\ \cite{Preis:2011sp}). The dashed line is the (second-order) 
onset of baryonic matter. The transition within the chirally restored phase between the ``LLL'' and ``hLL'' phases has disappeared because
baryonic matter is preferred in this region of the phase diagram.}
\label{fig:baryons}
\end{figure}
In order to discuss the inverse magnetic catalysis, we have plotted the zero-temperature phase diagram separately in 
Fig.\ \ref{fig:IMCSS}. This phase diagram shows intriguing similarities with the corresponding NJL phase diagram in Fig.\ \ref{Teq0pd}:
inverse magnetic catalysis is present at small magnetic fields and is most pronounced when the restored phase has a LLL--like behavior. Even the manifestation
of inverse magnetic catalysis in the analytical approximations is qualitatively the same as in the field-theoretical model as we now show. For large magnetic fields,
Eqs.\ (\ref{Omegabroken2}) and (\ref{OmegahLL}) can be used to write the free energy difference between restored and broken phases as 
\begin{equation} \label{ClogstonSS}
\Delta\Omega=\frac{N_cB}{4\pi^2}\left[\mu_q^2-M^2\frac{\sqrt{\pi}\Gamma\left(\frac{3}{5}\right)}{3\Gamma\left(\frac{1}{10}\right)}\right]
-\frac{N_cB}{16\pi^2}\mu_q^2 \, ,
\end{equation}
where we have identified $Ru_0/(2\pi\alpha')$ with the constituent quark mass $M$ \cite{Aharony:2006da,Johnson:2008vna}. 
This large-$B$ expression for $\Delta\Omega$ is remarkably 
similar to the weak-coupling expression (\ref{Clogston}) in the NJL model. We can thus conclude, for the reasons explained 
below Eq.\ (\ref{Clogston}), that in the large-$B$ regime the critical chemical potential must increase with $B$. This is confirmed by the 
chiral phase transition line of Fig.\ \ref{fig:IMCSS}. Note the difference between the 
terms $\propto \theta(\mu-M)$ in the NJL expression and the last term in Eq.\ (\ref{ClogstonSS}). Both terms come from a nonzero quark density which
in our NJL calculation is only present if $\mu>M$, while in our Sakai-Sugimoto calculation there is a topological quark density at nonzero 
$B$ for all $\mu$ due to the chiral spiral.  

For small magnetic fields we may apply an approximation in the spirit of Eq.\ (\ref{imc}). We compare the free energy of the broken phase
for small magnetic fields (\ref{Ombroken1}) with the free energy of the LLL phase (\ref{LLLcosts}). The result can be written as 
\begin{eqnarray} \label{imcSS}
\Delta\Omega\simeq-\frac{2N_c^{1/2}\Gamma\left(\frac{9}{16}\right)}{21\pi g_{\mathrm{YM}}\Gamma\left(\frac{1}{16}\right)}\,
M_\mathrm{KK}^{1/2}M_0^{7/2}+ \frac{N_c}{4\pi^2}B\mu_q^2-\frac{N_c^2g_{\rm YM}^2\eta_2(\mu)}{24\pi^3M_{\rm KK}R}\, B^2 \, , 
\end{eqnarray}
where $M_0\propto u_0^{(0)}$ is the constituent quark mass at $B=0$.
Again we recover the form of the NJL result (\ref{imc}). 
The main conclusion is that the energy cost for condensation is linear in $B$, whereas the energy gain from condensation, i.e., the magnetic 
catalysis is only quadratic in $B$ for small $B$. This allows for inverse magnetic catalysis. The dashed-dotted line in Fig.\ \ref{fig:IMCSS}
is the approximate phase transition from Eq.\ (\ref{imcSS}). Comparison with the full numerical result shows that the approximation captures 
the physics of inverse magnetic catalysis where it is most pronounced and that the ``hLL'' phase counteracts inverse magnetic catalysis. 

In Fig.\ \ref{fig:baryons} we show the phase diagram including baryonic matter discussed in \cite{Preis:2011sp}. The main observations
are that $(i)$ baryonic matter prevents chiral symmetry restoration for small magnetic field for any value of $\mu$ (as already found in 
Ref.\ \cite{Bergman:2007wp} for $B=0$) and that $(ii)$ for sufficiently
large magnetic fields, baryons become disfavored, i.e., the chirally broken, mesonic, phase is directly superseded by the quark matter phase.  
Interestingly, in the presence of baryonic matter, inverse magnetic catalysis becomes even more prominent in the phase diagram: 
now, the magnetic field restores chiral symmetry for any $\mu> 0.25$.

\section{Discussion}

We have investigated equilibrium phases at finite temperature, chemical potential, and magnetic field for one massless flavor in the 
Nambu--Jona-Lasinio model and the Sakai--Sugimoto model. 
For small flavor brane separations, the Sakai--Sugimoto model is conjectured to be dual to a (non-local) NJL model. 
Indeed, we have found intriguing qualitative similarities between both models. 

There is an exact equality of the number density at zero temperature of the lowest Landau level in the restored phase 
of the NJL model and the large magnetic field phase with restored chiral symmetry in the Sakai--Sugimoto model. The  
higher Landau level phase in the NJL model, however, differs from the small magnetic field phase with restored chiral symmetry in the 
Sakai--Sugimoto model. For example, there occur no de Haas--van Alphen oscillations in the holographic model. 
One possible interpretation is that in the holographic model -- dual to a strongly coupled gauge theory -- there are no quasiparticles and no 
sharp Fermi surface. Furthermore, the axial current found on the field theory side is also reproduced in the holographic model.
In the version of the model discussed here \cite{Bergman:2008qv}, the holographic current reproduces the field-theoretical current only 
up to a factor of $2$. 
This discrepancy can be resolved by properly implementing the axial anomaly \cite{Rebhan:2009vc}, however for the price of losing a consistent 
thermodynamic description. 

Also the phase diagrams in both models share the same qualitative features. The main differences are the order of the phase transitions 
(first and second order in NJL vs.\ first order in Sakai-Sugimoto), the saturation of the critical temperature and the critical chemical potential 
at asymptotically large magnetic fields (which only occurs in Sakai-Sugimoto), and the 
absence of de Haas--van Alphen oscillations of the phase transition line in the Sakai--Sugimoto model. 
The main physical effect, first discussed in detail in the holographic context \cite{Preis:2010cq}, is the nontrivial behavior of the 
chiral phase transition in a magnetic field at finite quark chemical potential. Somewhat unexpectedly, at sufficiently large chemical potentials 
and small temperatures and not too large magnetic fields, the effect of inverse magnetic catalysis dominates. We have explained inverse magnetic catalysis in both 
models by a free energy argument. This argument shows that, even if the magnetic field
increases the constituent quark mass (due to the usual magnetic catalysis) and thus increases the condensation energy, it also 
increases the energy cost for forming a chiral condensate. In particular, in the LLL, where the effect is most pronounced, the cost for overcoming
the separation of fermions and antifermions due to the chemical potential increases linearly in $B$, while the constituent quark mass rises 
quadratically. It is interesting that at asymptotically large magnetic fields the free energy difference in the Sakai-Sugimoto model 
resembles the corresponding expression in the weak-coupling limit of the NJL model. In this regime magnetic catalysis is dominant in both models, 
and the situation is analogous to weak-coupling superconductivity with mismatched Fermi surfaces. 

By fitting the parameters of the holographic model with the help of the critical temperature at $\mu=B=0$ from QCD lattice 
calculations \cite{Aoki:2006br,Aoki:2006we} and the (not very well known) critical chemical potential at $T=B=0$ from model calculations 
 \cite{Rebhan:2003wn,Kurkela:2009gj}, we find that inverse magnetic catalysis persists 
up to $B\simeq 1.0\times 10^{19}\ \mathrm{G}$, where the critical chemical potential has decreased from $400\ \mathrm{MeV}$ 
to about $230\ \mathrm{MeV}$. 
It is not clear whether the magnetic field inside compact stars is large enough to have any effect on the chiral phase transition. Our results
show, however, that if it is large enough then only {\it inverse} magnetic catalysis will play a role, i.e., the transition from 
hadronic to quark matter occurs at smaller densities than naively expected from the $B=0$ case.

We have included an anisotropic chiral condensate in the Sakai--Sugimoto model, but not in the NJL model. For comparison, it is easy to show that 
in the holographic calculation the assumption of an isotropic chiral condensate does not change the qualitative features of the phase diagram. 
One finds
that the effects of inverse magnetic catalysis are rather enhanced. On the other hand, including an anisotropic chiral condensate 
in the NJL model changes the phase diagram drastically \cite{Frolov:2010wn}. Most notably, there exists a phase with anisotropic chiral condensate 
even at $B=0$; in the Sakai--Sugimoto model, $B\neq0$ is necessary for having such a phase. Moreover, this phase inevitably has a 
finite quark density. In order to realize this in the holographic model at $B=0$ one needs solitonic 
baryon sources which are related to Skyrmions and thus rather different from ``baryons" in the NJL model which consist of dislocated quarks. 
We have briefly discussed the effect of such baryonic matter in the Sakai-Sugimoto model, based on Ref.\ \cite{Preis:2011sp}. One of the most 
important changes is the non-existence of a chiral symmetry restoration at $B=0$ for any value of the chemical potential.

Another phenomenon that was not included in our discussion is the so-called chiral shift \cite{Gorbar:2009bm,Gorbar:2011ya}, a chiral asymmetry 
in the Fermi 
surfaces of right- and left-handed charged fermions induced by a magnetic field. It would be interesting to discuss its effect on 
the chiral phase transition and thus on inverse magnetic catalysis. However, the chiral shift is related to the Fock exchange terms, 
which are suppressed at large $N_c$. Therefore, this effect
is difficult to study in a holographic model where $N_c\rightarrow \infty$ is necessary for the validity of the supergravity approximation.

\begin{acknowledgement}
This work has been supported by the Austrian science foundation FWF under project no.\ P22114-N16.
\end{acknowledgement}

\bibliographystyle{unsrt}
\bibliography{references}

\end{document}